\documentclass[12pt]{article}
\usepackage{graphicx,anysize,amsfonts,amsmath,amsthm,amssymb,enumerate,hyperref,color,setspace,booktabs,enumitem, subcaption}
\usepackage[font = normalsize]{caption}
\usepackage[mathscr]{euscript}
\usepackage[utf8]{inputenc}
\usepackage[english]{babel}
\usepackage{natbib}
\usepackage{moreverb, ragged2e}
\usepackage[font = normalsize]{subcaption}
\usepackage{rotating}
\newcommand\BibTeX{{\rmfamily B\kern-.05em \textsc{i\kern-.025em b}\kern-.08em
T\kern-.1667em\lower.7ex\hbox{E}\kern-.125emX}}
\newcommand\numberthis{\addtocounter{equation}{1}\tag{\theequation}}

\allowdisplaybreaks
\marginsize{1in}{1in}{1in}{1in}

\usepackage{sectsty}
\sectionfont{\fontsize{12}{15}\selectfont}
\subsectionfont{\fontsize{12}{15}\selectfont}

\usepackage{array}
\newcolumntype{L}[1]{>{\raggedright\let\newline\\\arraybackslash\hspace{0pt}}m{#1}}
\newcolumntype{C}[1]{>{\centering\let\newline\\\arraybackslash\hspace{0pt}}m{#1}}
\newcolumntype{R}[1]{>{\raggedleft\let\newline\\\arraybackslash\hspace{0pt}}m{#1}}
\providecommand{\keywords}[1]
{
  \small	
  \textbf{\textit{Keywords:}} #1
}

\title{\Large \textbf{Between- and Within-Cluster Spearman Rank Correlations}}
\author{\vspace{-1em}\small Shengxin Tu$^{1}$, 
Chun Li$^{2}$, and
Bryan E. Shepherd$^{1,*}$\\\vspace{-1em}
\small $^{1}$Department of Biostatistics, Vanderbilt University, Nashville, Tennessee, USA\\\vspace{-1em}
\small $^{2}$Department of Population and Public Health Sciences, University of Southern \small California,\\\vspace{-1em} \small Los Angeles, California, USA\\
\small $^{*}$bryan.shepherd@vanderbilt.edu}
\date{\vspace{-2ex}}

\begin{document}
\maketitle
\begin{abstract}
Clustered data are common in practice. Clustering arises when subjects are measured repeatedly, or subjects are nested in groups (e.g., households, schools). It is often of interest to evaluate the correlation between two variables with clustered data. There are three commonly used Pearson correlation coefficients (total, between-, and within-cluster), which together provide an enriched perspective of the correlation. However, these Pearson correlation coefficients are sensitive to extreme values and skewed distributions. They also vary with data transformation, which is arbitrary and often difficult to choose, and they are not applicable to ordered categorical data. Current nonparametric correlation measures for clustered data are only for the total correlation. Here we define population parameters for the between- and within-cluster Spearman rank correlations. The definitions are natural extensions of the Pearson between- and within-cluster correlations to the rank scale. We show that the total Spearman rank correlation approximates a linear combination of the between- and within-cluster Spearman rank correlations, where the weights are functions of rank intraclass correlations of the two random variables. We also discuss the equivalence between the within-cluster Spearman rank correlation and the covariate-adjusted partial Spearman rank correlation. Furthermore, we describe estimation and inference for the three Spearman rank correlations, conduct simulations to evaluate the performance of our estimators, and illustrate their use with data from a longitudinal biomarker study and a clustered randomized trial.  
\end{abstract}
\keywords{Clustered data; Rank association measures; Nonparametric correlation measures}
\clearpage

\pagenumbering{arabic} 

\section{Introduction}
\label{intro}
Clustered data are common in practice. Clustering arises when subjects (e.g., people) are measured repeatedly, or subjects are nested in clusters (e.g., households, schools) and measured only once. The total, between-, and within-cluster Pearson correlations are frequently used in the analysis of clustered data \citep{snijders1999, ferrari2005}. The total correlation measures the overall correlation but fails to acknowledge the clustered nature of the data. The between-cluster correlation measures the association between underlying variables at the cluster level, while the within-cluster correlation is the correlation after controlling for clustering. 

For example, in an observational study, people living with HIV on antiretroviral therapy (ART) had repeated measurements of CD4 and CD8 counts \citep{castilho2016}. There is interest in measuring the correlation between CD4 and CD8 counts subject to clustering. The between-cluster correlation measures the association between the underlying CD4 and CD8 counts in individuals. The within-cluster correlation describes the correlation between variations in CD4 and CD8 measures due to changes over time or measurement errors. Together, the total, between-, and within-cluster correlations provide a more complete picture of the relationship between CD4 and CD8 counts.

However, Pearson correlations are sensitive to extreme values and skewed distributions, and they depend on the scale of the data. For example, CD4 and CD8 counts are both right-skewed and sometimes transformed prior to analyses; estimates of the total, between-, and within-cluster Pearson correlations will vary with the transformation, which is arbitrary and often difficult to choose. Several measures may serve as alternatives to Pearson correlation, such as Spearman rank correlation, Kendall $\tau$, and copula-based measures of dependencies \citep{spearman1904, kendall1938, nelsen2006}.
Although Kendall $\tau$ is rank-based and bounded by $-1$ and $1$, it is a difference between concordance and discordance probabilities rather than a correlation coefficient. Copula-based measures of dependencies, initially developed for independent data and later extended to clustered data, require distributional assumptions and have complications with discrete data \citep{arakelian2014, kosmidis2016,genest2007}. In contrast, Spearman rank correlation is much more commonly used than the others and easily interpreted as a correlation coefficient. It can be empirically computed and easily applied to both continuous and discrete data. 

Some recent studies have proposed Spearman rank correlations for clustered data. Rosner and Glynn \citeyearpar{rosner2017} proposed a regression-based approach to obtain the maximum likelihood estimate of Pearson correlation for clustered data and then compute Spearman rank correlation by using its relationship with Pearson correlation under bivariate normality. Shih and Fay \citeyearpar{shih2017} defined Spearman rank correlation for clustered data as the Pearson correlation between the population versions of ridits \citep{Bross1958}, and applied within-cluster resampling and U-statistics for estimation and inference. Hunsberger, et al \citeyearpar{hunsberger2022} extended the work of Shih and Fay by improving the nominal level of the tests for clustered data with small sample sizes. However, these proposed rank correlations are only for the total correlation. There is a need to develop between- and within-cluster Spearman rank correlations. 

In this paper, we define population parameters of between- and within-cluster Spearman rank correlations, which are natural extensions of the traditional between- and within-cluster Pearson correlations to the rank scale. We show that the total Spearman rank correlation approximates a linear combination of the between- and within-cluster Spearman rank correlations, where the weights are functions of rank intraclass correlation coefficients \citep{tu2023}. We also show the equivalence between the within-cluster Spearman’s rank correlation and the covariate-adjusted partial Spearman’s rank correlation with clusters as covariates \citep{liu2018}. 

This paper is organized as follows. In Section 2, we briefly review Pearson correlations for clustered data. In Section 3, we introduce population parameters of the total, between-, and within-cluster Spearman rank correlations for clustered data. In Section 4, we illustrate the relationship between the three Spearman rank correlations. In Sections 5 and 6, we describe estimation and inference using semiparametric cumulative probability models. In Section 7, we conduct simulations to evaluate the performance of our estimators under different scenarios. In Section 8, we illustrate our method in two applications: a clustered randomized trial and a longitudinal biomarker study. Section 9 provides a discussion. Additional information is in the Supporting Information. We have developed an R package, \texttt{rankCorr}, available on CRAN, which implements our new method \citep{tuRankCorr}.

\section{Review of Pearson correlations for clustered data}
\label{pearson}
Let $(X, Y)^T$ denote a vector of two random variables from a two-level hierarchical joint distribution. It can be expressed as a sum,
\begin{align*}
    \begin{pmatrix} X \\ Y \end{pmatrix} = \begin{pmatrix} U_{X}\\ U_{Y} \end{pmatrix} + \begin{pmatrix} R_{X} \\ R_{Y}\end{pmatrix}, \numberthis \label{additive}
\end{align*} 
where $(U_{X}, U_{Y})^T$ is the random cluster mean (assuming it exists), and $(R_{X}, R_{Y})^T$ is the within-cluster deviation. The total correlation between $X$ and $Y$ is $\rho_t = corr(X,Y)$. The between-cluster correlation is the correlation between the cluster means, $\rho_b = corr(U_{X}, U_{Y})$. The within-cluster correlation is the correlation between the within-cluster deviations, $\rho_w = corr(R_{X}, R_{Y})$. 

The total correlation can be shown as a linear combination of the between- and within-cluster correlations, under the assumptions that $(U_{X}, U_{Y})^{T}\perp (R_{X}, R_{Y})^T$ and that the within-cluster covariance matrices are equal across clusters. Specifically, let the covariance matrix of $(U_{X}, U_{Y})^T$ be denoted as $ \begin{pmatrix} \sigma^2_{u} & \rho_b\sigma_{u}\eta_{u} \\   \rho_b\sigma_{u}\eta_{u} & \eta_{u}^2 \end{pmatrix}$ and the covariance matrix of $(R_{X}, R_{Y})^T$ be denoted as $\begin{pmatrix} \sigma^2_{r} & \rho_w\sigma_{r}\eta_{r} \\   \rho_w\sigma_{r}\eta_{r} & \eta_{r}^2 \end{pmatrix}$. The intraclass correlation coefficient (ICC) of $X$ is $\rho_{I_X}=\sigma^2_u/(\sigma^2_u+\sigma^2_r)$, which is the correlation between two random observations of $X$ in a random cluster. Similarly, the ICC of $Y$ is $\rho_{I_Y}=\eta^2_u/(\eta^2_u+\eta^2_r)$. Then we have 
\begin{align*} \rho_t & = \rho(X,Y)\\
& = \frac{\text{cov}(X,Y)}{\sqrt{\text{var}(X)\text{var}(Y)}} \\
& = \frac{\text{cov}(U_{X}, U_{Y}) + \text{cov}(R_{X}, R_{Y})}{\sqrt{\text{var}(X)\text{var}(Y)}} \\
& = \frac{\rho_b\sigma_u\eta_u + \rho_w\sigma_r\eta_r}{\sqrt{(\sigma^2_u + \sigma^2_r)(\eta^2_u + \eta^2_r)}} \\
& = \rho_{b}\sqrt{\rho_{I_X}\rho_{I_Y}}+\rho_{w}\sqrt{(1-\rho_{I_X})(1-\rho_{I_Y})}. \numberthis \label{pearsonrelation} \end{align*}

\section{Population parameters of Spearman rank correlations for clustered data}
\label{spearman}
Spearman rank correlation between two continuous variables is the correlation between their cumulative distribution functions (CDFs); this is known as the grade correlation \citep{kruskal1958}. More broadly, for any type of orderable variables, it is the correlation between population versions of midranks or ridits \citep{kendall1970, Bross1958}. Given any CDF $F$, let $F(x-)=\lim_{t \uparrow x} F(t)$ and $F^*(x) = \big\{F(x) + F(x-)\big\}/2$. If the distribution is continuous, then $F^*(x) = F(x)$. If the distribution is discrete or mixed, $F^*(x)$ corresponds to the population versions of ridits \citep{Bross1958}. The population parameter of Spearman rank correlation between two random variables $X$ and $Y$ with CDFs $F_X$ and $F_Y$ is denoted as $\gamma(X, Y) = corr\big\{F^*_X(X), F^*_Y(Y)\big\}$ \citep{kendall1970, liu2018}.

Let $X$ and $Y$ denote two random variables from a two-level hierarchical joint distribution, with a random variable $Z$ indicating cluster. Henceforth, we do not assume the additive model ($\ref{additive}$). The total Spearman rank correlation is the overall rank correlation between $X$ and $Y$. We define its population parameter as
\begin{align*}
\gamma_t  = corr\big\{F^*_{X}(X), F_{Y}^*(Y)\big\}.
 \numberthis \label{total}
\end{align*}
With continuous $X$ and $Y$, $\gamma_t = 12cov\big\{F_{X}(X), F_{Y}(Y)\big\}$, because $F^*_X(X) = F_X(X) \sim Unif(0,1)$, $F^*_Y(Y) = F_Y(Y) \sim Unif(0,1)$, and their variances equal $1/12$.

Let $F_{X |Z}$ and $F_{Y | Z}$ be the CDFs of $X$ and $Y$ conditional on $Z$, respectively. The population parameter of the within-cluster Spearman rank correlation is defined as 
\begin{align*}
\gamma_w  = corr\big\{F^*_{X | Z}(X), F^*_{Y|Z}(Y)\big\}.
 \numberthis \label{within}
\end{align*}
Note that $\gamma_w$ is not a function of cluster index and that it does not assume an equal variance structure across clusters. In fact, $\gamma_w$ is identical to the covariate-adjusted partial Spearman rank correlation \citep{liu2018}, where the covariates are the clusters, $Z$. Since the partial Spearman rank correlation can be expressed using probability-scale residuals (PSRs) \citep{li2012, shepherd2016}, we can express $\gamma_w$ similarly. The PSRs of $X=x$ and $Y=y$ given $Z$ are defined as $r(x,F_{X|Z}) = 2F^*_{X|Z}(x)-1$ and $r(y,F_{Y|Z})=2F^*_{Y|Z}(y)-1$, respectively. Then we have 
$$\gamma_{w}=corr\Big\{r\big(X,F_{X|Z}\big), r\big(Y,F_{Y|Z}\big)\Big\}.$$
This connection allows us to derive an estimator for $\gamma_w$, which will be described in Section \ref{estimation}. 

The usage of cluster means is not desirable for the between-cluster Spearman rank correlation because means are scale-dependent and sensitive to outliers and skewness. We use the general concept of cluster centroids to define the between-cluster Spearman rank correlation. A cluster centroid defines the central tendency of random variables in the same cluster. It is usually the median. Let $\tilde X$ and $\tilde Y$ be the cluster centroids, and let $F_{\tilde X}$ and $F_{\tilde Y}$ be their CDFs, respectively. Assuming that clusters are independent, the between-cluster Spearman rank correlation treats clusters as units of interest and measures the association between cluster centroids. We define its population parameter as 
\begin{align*}
\gamma_b  = corr\big\{F^*_{\tilde X}(\tilde X), F_{\tilde Y}^*(\tilde Y)\big\}.
 \numberthis \label{between}
\end{align*}

Our definitions of $\gamma_t$, $\gamma_b$, and $\gamma_w$ are easily interpreted as rank correlations. In the special case where $(X, Y)^T$ has a similar hierarchical population model as (\ref{additive}) in Section \ref{pearson} except that $(U_{X}, U_{Y})^T$ is the cluster median and $(R_{X}, R_{Y})^T$ has a median of zero, then $\gamma_t = \gamma(X, Y)$, $\gamma_b = \gamma(U_{X}, U_{Y})$, and $\gamma_w = \gamma(R_{X}, R_{Y})$. 

Furthermore, our definitions of $\gamma_t$, $\gamma_b$, and $\gamma_w$ are also applicable to ordered categorical data. While the definitions of $\gamma_t$ and $\gamma_w$ in (\ref{total}) and (\ref{within}) can be directly applied, the definition of $\gamma_b$ in (\ref{between}) needs an extension. For an ordered categorical variable $X$, the median is defined as any category $c$ for which $P(X \leq c) \geq 0.5$ and $P(X \geq c) \geq 0.5$. The median is often a unique value. In the rare situation where $P(X \leq c) = 0.5$, both the category $c$ and the next higher category (denoted as $c+$) are the medians, and we define the cluster centroid $\tilde X$ as $c$ with a probability of 0.5 and $c+$ with a probability of 0.5. If there are clusters like this with two cluster medians for a variable, we define $\gamma_b=E\big[\gamma(\tilde X, \tilde Y)\big]$, the expectation of the Spearman rank correlation over all possible combinations of cluster medians in the population. If no clusters have two cluster medians, the definition in (\ref{between}) can be directly applied.  

\section{Relationship between \texorpdfstring{\MakeLowercase{$\gamma_t$, }}, \texorpdfstring{\MakeLowercase{$\gamma_w$, }}, and \texorpdfstring{\MakeLowercase{$\gamma_b$}}{}}
\label{relationship}
The total Spearman rank correlation can be decomposed into two weighted components. The weights are functions of the rank ICC, which is a natural extension of Fisher's ICC \citep{fisher1925} to the rank scale \citep{tu2023}. Let $(X, X^{'})$ be a random pair from a random cluster. The rank ICC of $X$ is 
\begin{align*}
\gamma_{I_X} & =  corr\big[F_{X}^*( X_{}), F_{X}^*(X^{'})\big]\\
& = cov\big[F^*_{X}(X_{}), F^*_{X}(X^{'})\big] / var\big[F^*_{X}( X_{})\big] \\
& = cov\Big\{E\big[F^*_{ X}( X_{})| Z \big], E\big[F^*_{ X}( X^{'})| Z\big]\Big\} / var\big[F^*_{ X}( X_{})\big] + E\Big\{cov\big[F^*_{ X}( X_{}), F^*_{ X}( X^{'}) | Z\big]\Big\} / var\big[F^*_{ X}( X_{})\big]\\
& = var\Big\{E\big[F^*_{ X}( X)| Z\big]\Big\} / var\big[F^*_{ X}( X)\big] + E\Big\{cov\big[F^*_{ X}( X_{}), F^*_{ X}( X^{'}) | Z\big]\Big\} / var\big[F^*_{ X}( X_{})\big] \\
& = var\Big\{E\big[F^*_{ X}( X)| Z\big]\Big\} / var\big[F^*_{ X}( X)\big] + D_X,
\end{align*}
where $D_X=E\Big\{cov\big[F^*_{ X}( X), F^*_{ X}( X^{'}) |  Z\big]\Big\} / var\big[F^*_{ X}( X)\big]$. Note that line 3 equals line 4 because $E\big[F_{ X}^*( X_{})| Z\big] =
E\big[F^*_{X}( X^{'})| Z\big]$. When cluster sizes in the population are finite, $D_X$ is negative. When cluster sizes in the population are infinite, $D_X$ is equal to 0. The rank ICC of $Y$, $\gamma_{I_Y}$, is similarly defined. 

The decomposition of the total Spearman rank correlation is 
\begin{align*}
 \gamma_{t}  & = corr\big\{F_{ X}^*( X_{}), F^*_{ Y}( Y_{})\big\} \\
 & = \frac{cov\big\{F_{ X}^*( X_{}), F^*_{ Y}( Y_{})\big\}}{\sqrt{var\big[F_{ X}^*( X_{})\big] var\big[ F^*_{ Y}( Y_{})\big]}} \\
& = \frac{cov\Big\{E\big[F_{ X}^*( X_{})| Z\big],E\big[F^*_{ Y}( Y_{})  | Z\big]\Big\} + E\Big\{cov\big[F_{ X}^*( X_{}), F^*_{ Y}( Y_{})  | Z\big]\Big\}}{\sqrt{var\big[F_{ X}^*( X_{})\big] var\big[ F^*_{ Y}( Y_{})\big]}}\\
& = \frac{cov\Big\{E\big[F_{ X}^*( X_{})| Z\big],E\big[F^*_{ Y}( Y_{})  | Z\big]\Big\}}{var\Big\{E\big[F_{ X}^*( X_{})| Z\big]\Big\}var\Big\{E\big[F^*_{ Y}( Y_{})  | Z\big]\Big\}} \sqrt{\frac{var\Big\{E\big[F^*_{ X}( X_{})| Z\big]\Big\}}{var\big[F^*_{ X}( X_{})\big]}}\sqrt{\frac{var\Big\{E\big[F^*_{Y}(Y_{})| Z\big]\Big\}}{var\big[F^*_{Y}(Y_{})\big]}}\\
& + \frac{E\Big\{cov\big[F_{ X}^*( X_{}), F^*_{ Y}( Y_{})  | Z\big]\Big\}} {\sqrt{E\Big\{var\big[F_{ X}^*( X_{})|Z\big]\}E\Big\{var\big[F_{ Y}^*( Y_{})|Z\big]\Big\}}} \sqrt{\frac{E\Big\{var\big[F_{ X}^*( X_{})|Z\big]\Big\}}{var\big[F^*_{ X}( X_{})\big]}}\sqrt{\frac{E\Big\{var\big[F_{ Y}^*( Y_{})|Z\big]\Big\}}{var\big[F^*_{Y}(Y_{})\big]}} \\ 
& =  corr\Big\{E\big[F_{ X}^*( X_{})| Z\big],E\big[F^*_{ Y}( Y_{})  | Z\big]\Big\}\sqrt{(\gamma_{I_X} - D_{X})(\gamma_{I_Y} - D_{Y})} \\
& + \frac{E\Big\{cov\big[F_{ X}^*( X_{}), F^*_{ Y}( Y_{})  | Z\big]\Big\}} {\sqrt{E\Big\{var\big[F_{ X}^*( X_{})|Z\big]\Big\}E\Big\{var\big[F_{ Y}^*( Y_{})|Z\big]\Big\}}} \sqrt{(1-\gamma_{I_X} + D_{X})(1-\gamma_{I_Y} + D_{Y})} \\
& = S_1 \sqrt{(\gamma_{I_X} - D_{X})(\gamma_{I_Y} - D_{Y})} + S_2  \sqrt{(1-\gamma_{I_X} + D_{X})(1-\gamma_{I_Y} + D_{Y})},
\end{align*}
where $S_1=corr\Big\{E\big[F_{ X}^*( X)| Z\big],E\big[F^*_{ Y}( Y)  | Z\big]\Big\}$ and $S_2=\frac{E\Big\{cov\big[F_{ X}^*( X), F^*_{ Y}( Y)  | Z\big]\Big\}} {\sqrt{E\Big\{var\big[F_{ X}^*( X)|Z\big]\Big\}E\Big\{var\big[F_{ Y}^*( Y)|Z\big]\Big\}}}$. When the cluster size in the population is infinite, then $D_X = D_Y=0$,
$$ \gamma_{t} = S_1\sqrt{\gamma_{I_X}\gamma_{I_Y}} + S_2 \sqrt{(1-\gamma_{I_X})(1-\gamma_{I_Y})}.$$
Simulations suggest that $S_1$ and $S_2$ can be approximated by $\gamma_b$ and $\gamma_w$ in general, respectively (Web Appendix B). That is, 
\begin{align*}
     \gamma_{t} \approx \gamma_b\sqrt{(\gamma_{I_X} - D_{X})(\gamma_{I_Y} - D_{Y})} + \gamma_w \sqrt{(1-\gamma_{I_X} + D_{X})(1-\gamma_{I_Y} + D_{Y})}. \numberthis \label{spearmanrelation}
\end{align*}
If the cluster size in the population is large, then $D_X \approx D_Y \approx 0$ and we have 
\begin{align*}
    \gamma_{t} \approx \gamma_b\sqrt{\gamma_{I_X}\gamma_{I_Y}} + \gamma_w\sqrt{(1-\gamma_{I_X})(1-\gamma_{I_Y})}. \numberthis \label{spearmanrelation2}
\end{align*}
This relationship is similar to that for Pearson correlations in (\ref{pearsonrelation}), which was derived for the additive model (\ref{additive}) with infinite cluster sizes.  

\begin{figure}
     \centering
     \begin{subfigure}[b]{0.5\textwidth}
         \centering
         \includegraphics[width=0.75\textwidth]{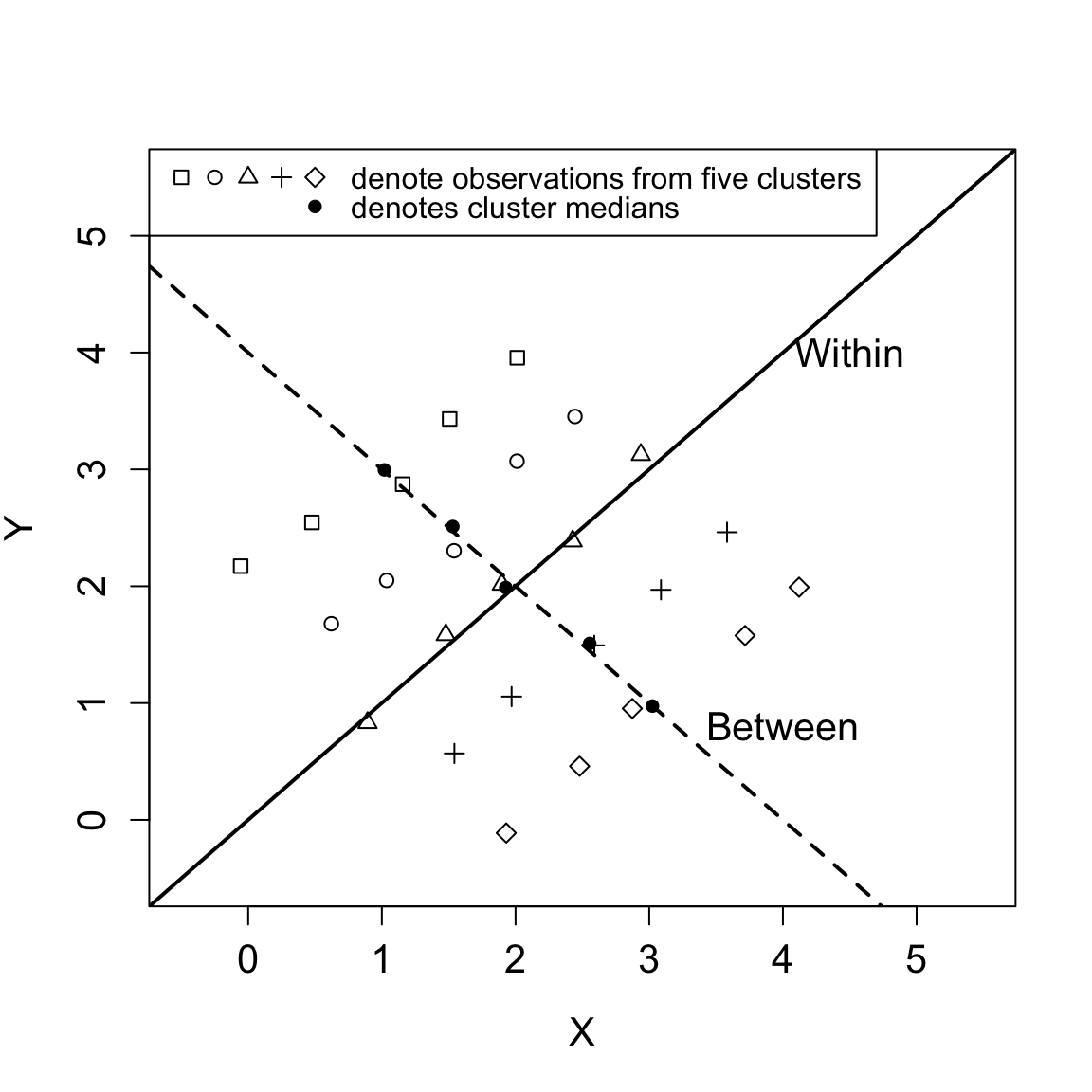} \vspace{-0.5cm}
         \caption{$(\gamma_t,\gamma_b,\gamma_w,\gamma_{I_X}, \gamma_{I_Y})=(0, -1, 1, 0.5, 0.5)$}
         \label{fig:fig1.1}
     \end{subfigure}\hfill
     \begin{subfigure}[b]{0.5\textwidth}
         \centering
         \includegraphics[width=0.75\textwidth]{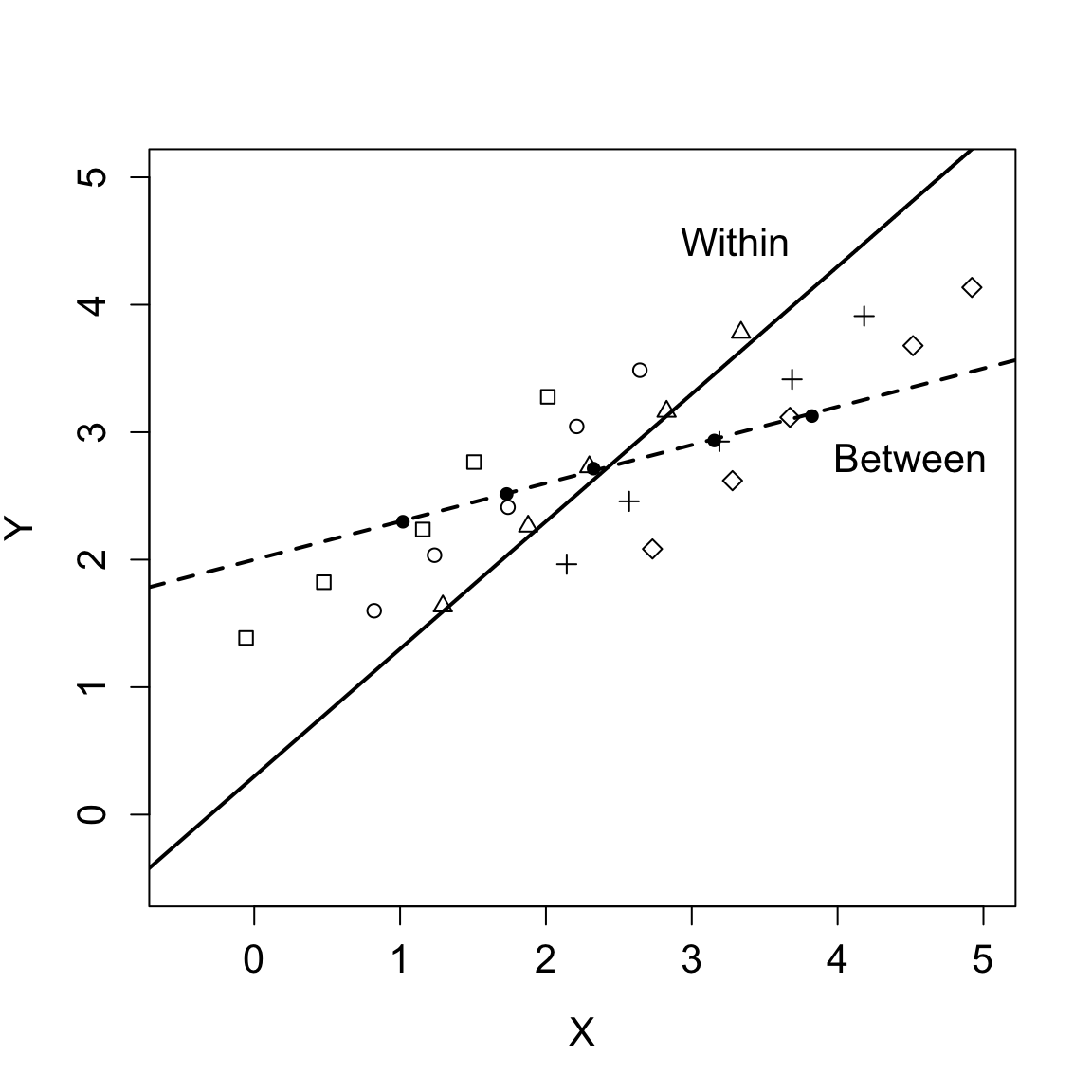}
         \vspace{-0.5cm}
         \caption{$(\gamma_t,\gamma_b,\gamma_w,\gamma_{I_X}, \gamma_{I_Y})=(0.84, 1, 1, 0.6, 0.1)$}
         \label{fig:fig1.2}
     \end{subfigure}\\
          \begin{subfigure}[b]{0.5\textwidth}
         \centering
         \includegraphics[width=0.75\textwidth]{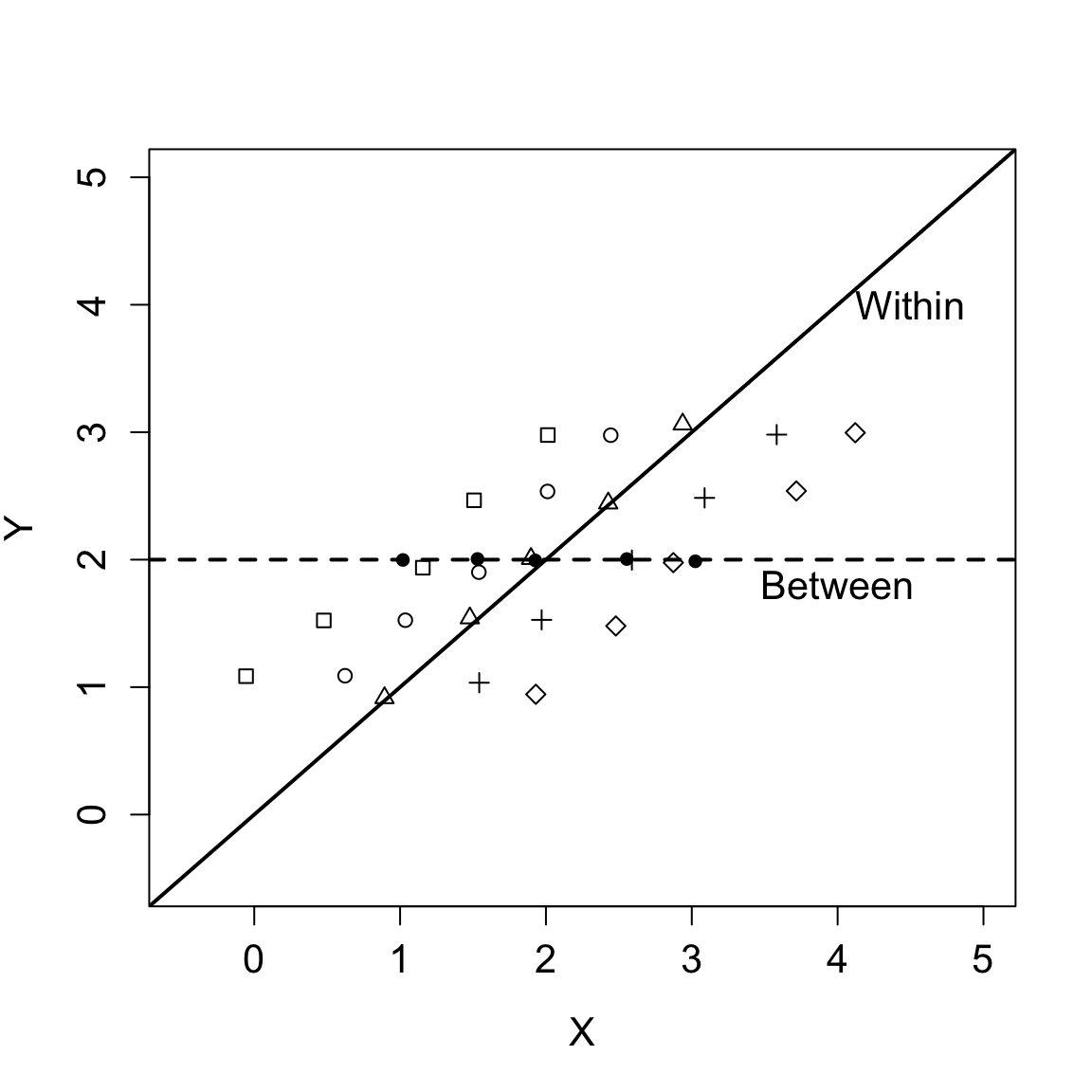}
         \vspace{-0.5cm}
         \caption{$(\gamma_t,\gamma_b,\gamma_w,\gamma_{I_X}, \gamma_{I_Y})=(0.71, 0, 1, 0.5, 0)$}
         \label{fig:fig1.3}
     \end{subfigure}\hfill
     \begin{subfigure}[b]{0.5\textwidth}
         \centering
         \includegraphics[width=0.75\textwidth]{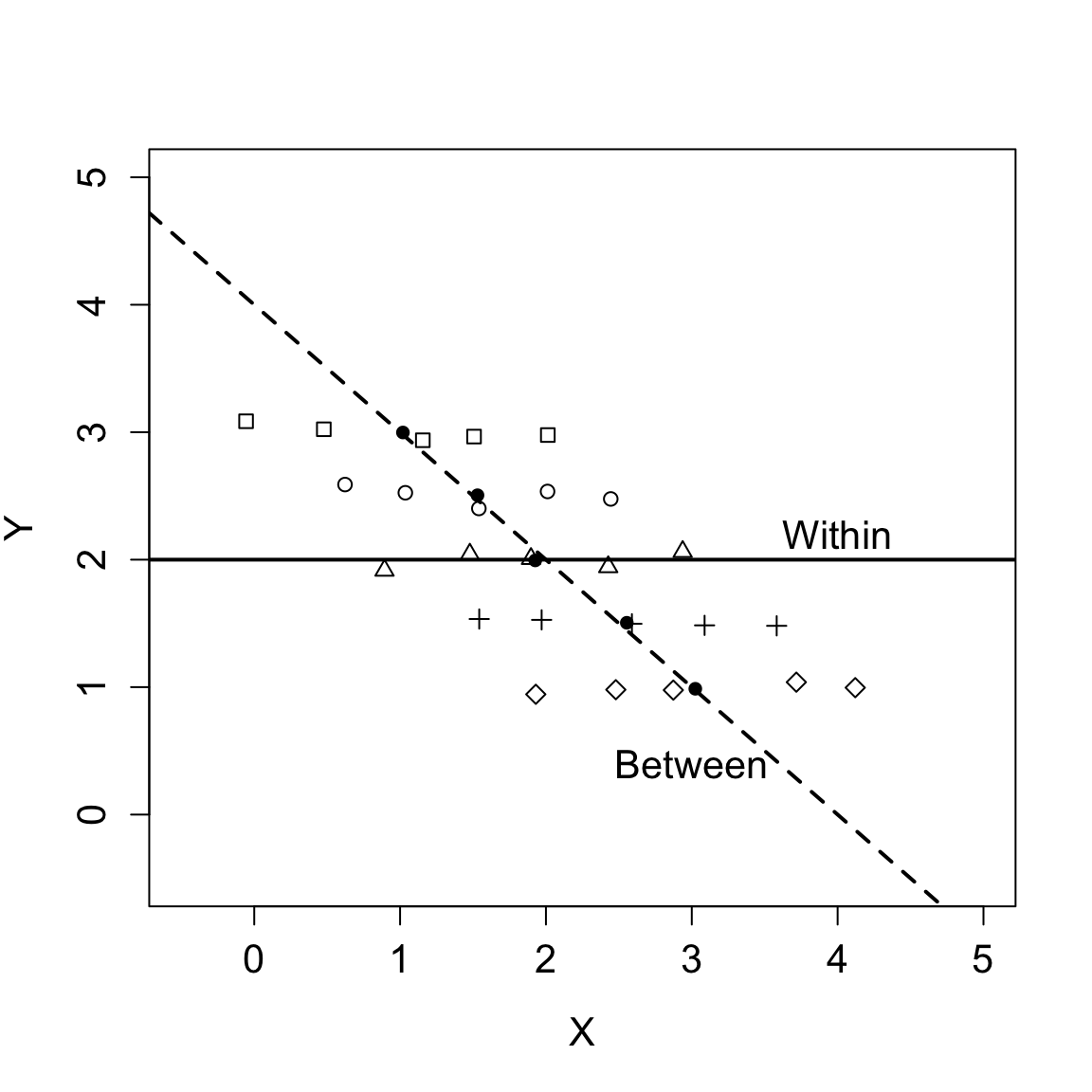}
         \vspace{-0.5cm}
         \caption{$(\gamma_t,\gamma_b,\gamma_w,\gamma_{I_X}, \gamma_{I_Y})=(-0.71, -1, 0, 0.5, 1)$}
         \label{fig:fig1.4}
     \end{subfigure}\\
    \begin{subfigure}[b]{0.5\textwidth}
         \centering
         \includegraphics[width=0.75\textwidth]{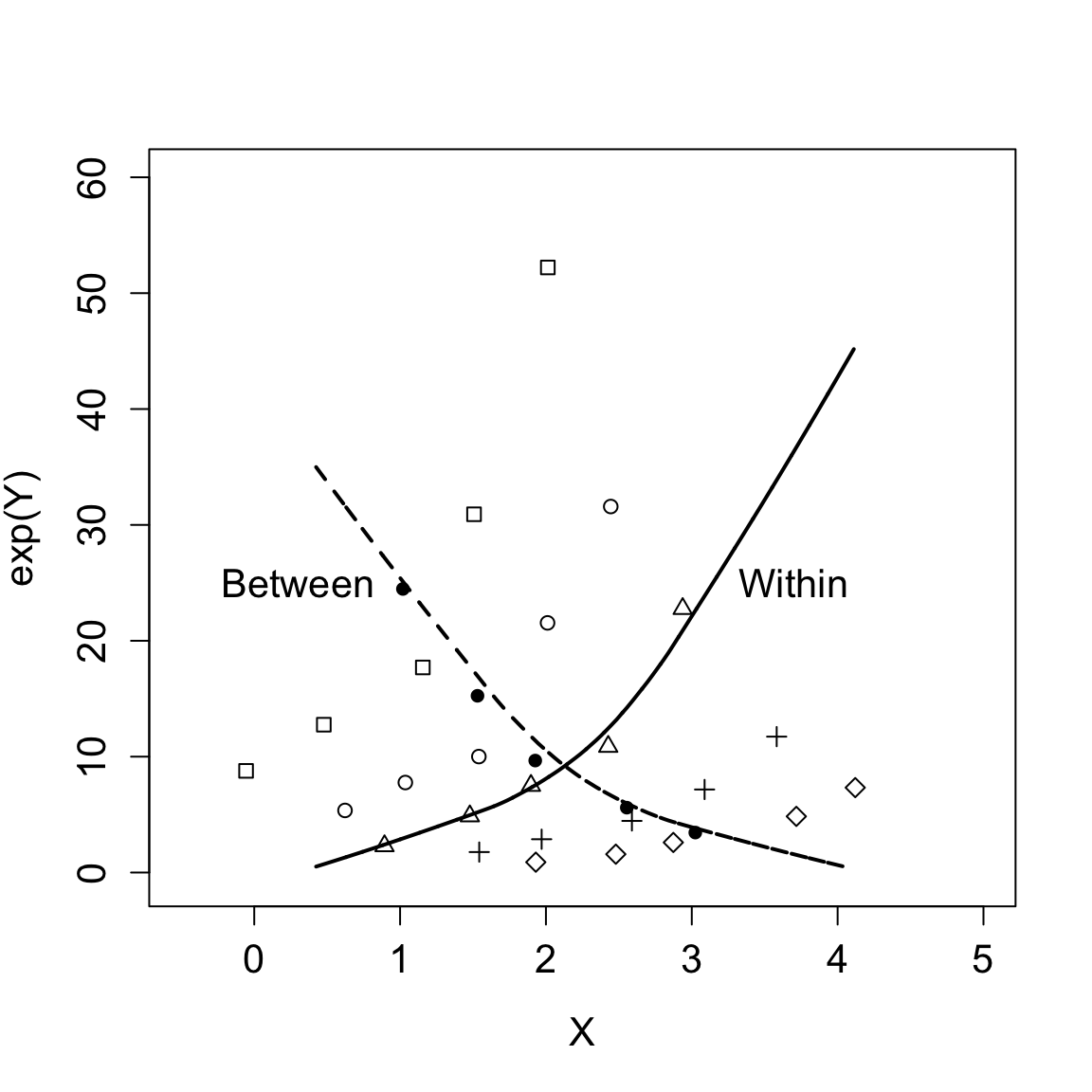}
         \vspace{-0.5cm}
         \caption{$(\gamma_t,\gamma_b,\gamma_w,\gamma_{I_X}, \gamma_{I_Y})=(0, -1, 1, 0.5, 0.5)$}
         \label{fig:fig1.5}
     \end{subfigure}\hfill
     \begin{subfigure}[b]{0.5\textwidth}
         \centering
         \includegraphics[width=0.75\textwidth]{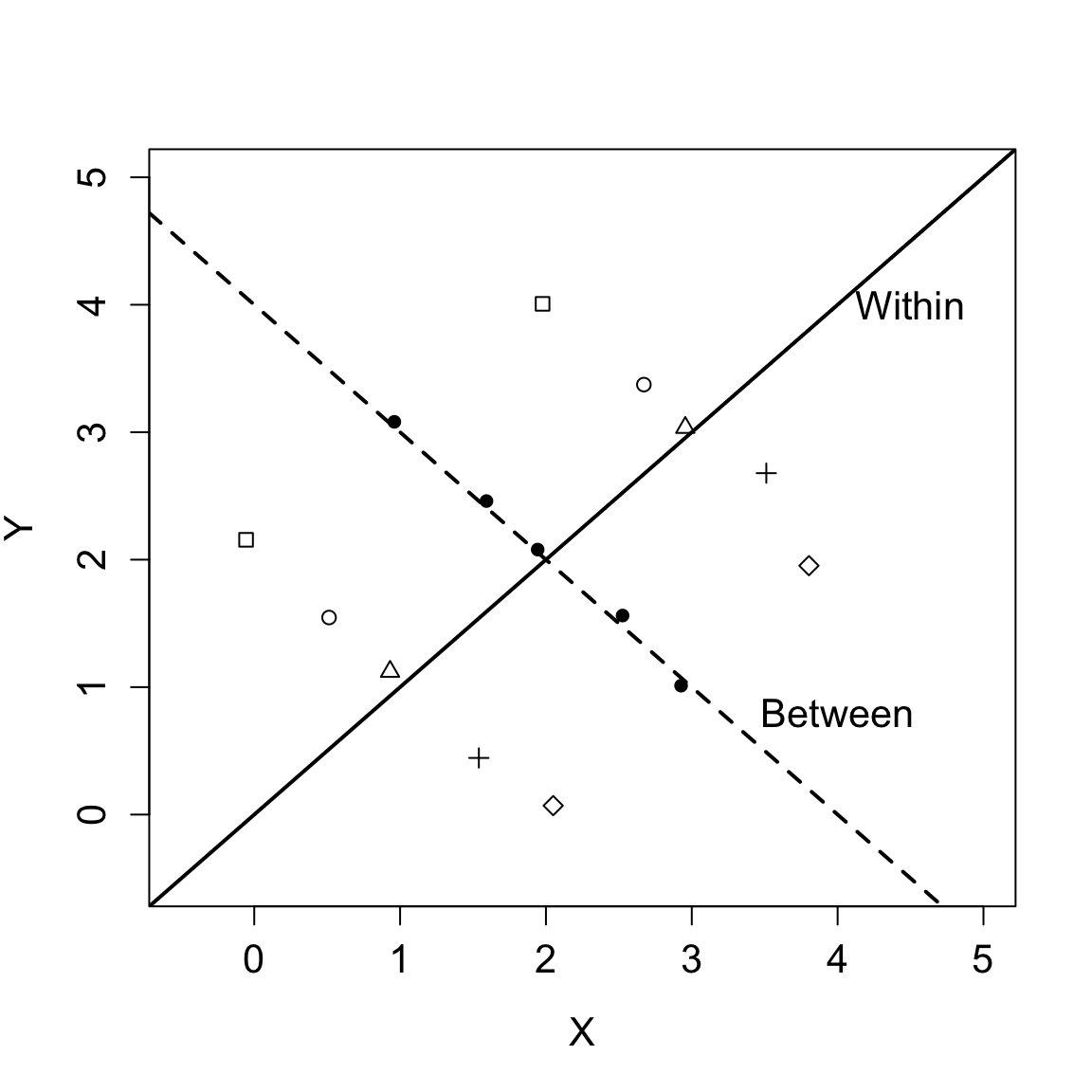}
         \vspace{-0.5cm}
         \caption{$(\gamma_t,\gamma_b,\gamma_w,\gamma_{I_X}, \gamma_{I_Y})=(0, -1, 1, -1, -1)$}
         \label{fig:fig1.6}
     \end{subfigure}\\
     \caption[]{Toy examples for the relationship between total ($\gamma_t$), between-cluster ($\gamma_b$), within-cluster ($\gamma_w$) Spearman rank correlations and the rank intraclass correlations ($\gamma_{I_X}$, $\gamma_{I_Y}$). For illustration purposes, we show five clusters and five (two for (f)) observations to represent the distribution of each cluster. Black dots represent cluster medians, and the other symbols represent the five clusters. The solid lines show the direction of the within-cluster correlation and the dashed lines show the direction of the between-cluster correlation.}
        \label{fig:fig1}
\end{figure}

We provide some toy examples to illustrate the relationship between $\gamma_t$, $\gamma_b$, and $\gamma_w$ under different rank ICCs of $X$ and $Y$ (Figure \ref{fig:fig1}). Figures 1a and 1b show examples where $\gamma_b$ and $\gamma_w$ are in the opposite or same directions, respectively. If $X$ and $Y$ have moderate rank ICCs of 0.5, then $\gamma_b$ and $\gamma_w$ contribute equally to $\gamma_t$, and $\gamma_t$ is the average of $\gamma_b$ and $\gamma_w$ (Figures 1a and 1e). If the rank ICCs are large, $\gamma_t$ is dominated by $\gamma_b$, while if the rank ICCs are low, $\gamma_t$ is dominated by $\gamma_w$. More extremely, if one of the rank ICCs is close to 1, $\gamma_t$ is close to $\gamma_b\sqrt{\gamma_{I_X}\gamma_{I_Y}}$ (Figure 1d). On the contrary, if one of the rank ICCs is near 0, which means that the observations in a cluster are nearly independent, then $\gamma_t$ is close to $\gamma_w\sqrt{(1-\gamma_{I_X})(1-\gamma_{I_Y})}$ (Figure 1c). When cluster sizes in the population are finite, the rank ICCs can be negative.  If any of the rank ICCs is negative, the relationship between $\gamma_t$, $\gamma_b$, and $\gamma_w$ is (\ref{spearmanrelation}) rather than the simpler (\ref{spearmanrelation2}). Figure 1f illustrates an extreme example where the rank ICCs are both $-1$. (This happens when cluster sizes are two.) In this example, $\gamma_b$ and $\gamma_w$ are strong and opposite whereas the total correlation is zero.

\section{Estimation}
\label{estimation}
Since the total Spearman rank correlation is the Pearson correlation between $F^*_X$ and $F^*_Y$, our estimator of $\gamma_t$ is $\hat \gamma_t = corr(\hat F_X^*, \hat F_Y^*)$, which is a plug-in estimator. Given two-level data $\{(x_{ij}, y_{ij}): i=1,2,...,n, j=1,2,...,k_i\}$ with a total number of observations of $N = \sum_{i=1}^{n} k_i$, where $i$ denotes the cluster and $j$ denotes the index within a cluster. A nonparametric estimator of the CDF of $X$ is $\hat F_X(x) = \sum_{i=1}^n \sum_{j=1}^{k_i} w_{ij}I(x_{ij} \leq x)$, where $w_{ij}$ is the weight of observation $x_{ij}$ and $\sum_{i=1}^n \sum_{j=1}^{k_i} w_{ij} = 1$. The weight $w_{ij}$ depends on how we believe the data reflect the composition of the underlying hierarchical distribution; for example, $w_{ij}=1/(nk_i)$ corresponds to equal weights for clusters and $w_{ij}=1/N$ corresponds to equal weights for observations \citep{tu2023} Similarly, we estimate $F_X(x-)$ with $\hat F_X(x-) = \sum_{i=1}^n \sum_{j=1}^{k_i} w_{ij}I(x_{ij} < x)$, and define $\hat F^*_X(x) = \big\{\hat F_X(x) + \hat F_X(x-)\big\}/2$. Estimation of $F^*_Y$ is similar. The general form of our estimator of $\gamma_t$ is
$$\hat \gamma_t = \frac{\sum\limits_{i=1}^n\sum\limits_{j=1}^{k_i} w_{ij} \Big[\hat F^*_X(x_{ij}) -  \bar{\hat{F^*_X}}\Big]\Big[\hat F^*_Y(y_{ij}) -  \bar{\hat{F^*_Y}}\Big]}{\sqrt{\sum\limits_{i=1}^n\sum\limits_{j=1}^{k_i} w_{ij} \Big[\hat F^*_X(x_{ij}) -  \bar{\hat{F^*_X}}\Big]^2}\sqrt{\sum\limits_{i=1}^n\sum\limits_{j=1}^{k_i} w_{ij} \Big[\hat F^*_Y(y_{ij}) -  \bar{\hat{F^*_Y}}\Big]^2}},$$
where $\bar{\hat{F^*_X}} = \sum\limits_{i=1}^n\sum\limits_{j=1}^{k_i} w_{ij} \hat F^*_X(x_{ij})$
and $\bar{\hat{F^*_Y}} = \sum\limits_{i=1}^n\sum\limits_{j=1}^{k_i} w_{ij} \hat F^*_Y(y_{ij})$. If we assign equal weights to clusters (i.e., $w_{ij}=1/(nk_i)$), our estimator of the total Spearman rank correlation is equal to the estimator of Shih and Fay \citep{shih2017}. 

We show in Section \ref{spearman} that $\gamma_w$ is identical to the covariate-adjusted partial Spearman rank correlation and can be expressed in terms of PSRs, suggesting that $\gamma_w$ can be estimated by sample PSRs \citep{liu2017}. Hence, our estimator of $\gamma_w$ is
\begin{equation}
  \hat \gamma_w = \frac{\sum\limits_{i=1}^n\sum\limits_{j=1}^{k_i} w_{ij} (x_{ij,res} -  \bar{x}_{res})(y_{ij,res} -  \bar{y}_{res})}{\sqrt{\sum\limits_{i=1}^n\sum\limits_{j=1}^{k_i} w_{ij} (x_{ij,res} - \bar{x}_{res})^2}\sqrt{\sum\limits_{i=1}^n\sum\limits_{j=1}^{k_i} w_{ij} (y_{ij,res} - \bar{y}_{res})^2}}, \label{est_gamma} 
\end{equation}
where $x_{ij,res} = r(x_{ij}, \hat F_{X|Z})$, $y_{ij,res} = r(y_{ij}, \hat F_{Y|Z})$, $\bar{x}_{res} = \sum\limits_{i=1}^n\sum\limits_{j=1}^{k_i} w_{ij} x_{ij,res}$, $\bar{y}_{res} = \sum\limits_{i=1}^n\sum\limits_{j=1}^{k_i} w_{ij} y_{ij,res}$. We can obtain PSRs using nonparametric, parametric, or semiparametric models. A nonparametric estimator of $F_{X|Z}$ is simply the empirical CDF of $X$ conditional on $Z$. Estimators from nonparametric models are the most robust but can be inefficient and unstable if cluster sizes are small. Parametric models are the most efficient under correct assumptions but less robust to extreme values, sensitive to model misspecification or outcome transformation, and not congruent with the spirit of Spearman rank correlation. To achieve a compromise between robustness and efficiency, we employ semiparametric models in which only the order information of outcomes is used and the clusters share a common latent variable distribution except for cluster-specific shifts. This way we can borrow information across clusters and still maintain the rank-based nature of Spearman rank correlation. 

Specifically, we designate cluster 1 as the reference cluster, and define $Z_i$ ($i=2,\ldots,n$) as an indicator variable such that $Z_i=1$ when the observation is in cluster $i$ and $Z_i=0$ otherwise. We then model $X$ and $Y$ on $Z=(Z_2,\ldots,Z_n)^T$ to obtain PSRs for $X$ and $Y$, respectively. Here we incorporate the semiparametric linear transformation model, which assumes a linear relationship between the outcome $X$ and the covariates $Z$ after an unspecified monotonic transformation that is empirically estimated. Specifically, $X^* = \alpha_X(X) = \beta_X^T Z + \epsilon$, where $X^*$ is a latent variable, $\alpha_X(\cdot)$ is the unspecified transformation, $\beta_X = (\beta_{X2},\ldots,\beta_{Xn})^T$, and $\epsilon$ follows a known distribution \citep{liu2017} This model is equivalently written as the cumulative probability model (CPM), $g_X\big\{F_{X|Z}(x)\big\} = \alpha_X(x) - \beta_X^T Z$, where $g_X(\cdot) = F^{-1}_{\epsilon}$ is a link function \citep{liu2017} Model estimation can be implemented using software for fitting ordinal cumulative probability (``link'') models with each unique outcome representing a separate ordinal category and $\alpha_X(\cdot)$ estimated with a step function. For example, the \texttt{orm()} function in the \texttt{rms} package of R can be used \citep{harrell2015} A similar model is fit for $Y$ on $Z$. After obtaining PSRs from the CPMs of $X$ on $Z$ and of $Y$ on $Z$, we then simply estimate $\gamma_{w}$ as in (\ref{est_gamma}).

As mentioned in Section \ref{spearman}, we often use the cluster median as the cluster centroid, so $\gamma_b$ is Spearman rank correlation between cluster medians. One simple estimation approach is to estimate $\gamma_b$ as Spearman rank correlation between the sample cluster medians (i.e., $\{(\hat{\tilde{x_{i}}}, \hat{\tilde{y_{i}}}): i=1,2,...,n\}$, where $\hat{\tilde{x_{i}}}$ and $\hat{\tilde{y_{i}}}$ are the medians of $\{x_{i1},...,x_{ik_i}\}$ and $\{y_{i1},...,y_{ik_i}\}$, respectively). However, this approach only uses information within clusters to estimate cluster medians, which can have high variation with small cluster sizes. Thus, we consider estimating the cluster medians using CPMs. The CPMs borrow information across clusters and their estimates of cluster medians are less variable than the simple estimates. Moreover, for ordered categorical data, the CPMs allow us to obtain cluster medians on the latent variable scale, thus simplifying the estimation of $\gamma_b$ by eliminating the need to consider all possible combinations of cluster medians on the original scale in the presence of clusters with two medians.

Let us consider a CPM of $X$ on $Z$, $g_X\{F_{X|Z}(x)\} =\alpha_X(x)-\beta_X^TZ$. For any $Z=z$, let $x_z$ be the true median of $X$ given $Z=z$. Since $F_{X|Z=z}(x_z)=0.5$ and $g_X(0.5)=c$ is a constant ($c=0$ for symmetric link functions), then we have $c =\alpha_X(x_z)-\beta_X^Tz$. That is, the monotone function $\alpha_X$ transforms the median $x_z$ to $\beta_X^Tz + c$. In the setting of clustered data, $Z=(Z_2,\ldots,Z_n)^T$ is a vector of indicator variables for the clusters, and thus the cluster medians are $c$ for cluster 1 and $\beta_{Xi}+c$ for cluster $i$ ($i=2,\ldots,n$).  Since $\alpha_X$ is a monotonic increasing transformation, a Spearman rank correlation that involves the cluster medians of $X$ can be computed with $(0,\beta_{X2},\ldots,\beta_{Xn})^T + c$ or equivalently $(0,\beta_{X2},\ldots,\beta_{Xn})^T$. Similarly, a Spearman rank correlation that involves the cluster medians of $Y$ can be computed with $(0,\beta_{Y2},\ldots,\beta_{Yn})^T$.  All these values can be estimated from the CPMs. Thus, our estimator of $\gamma_b$ is the rank correlation over the $n$ pairs of estimated cluster medians, $\{(0,0), (\hat\beta_{X2},\hat\beta_{Y2}), \ldots, (\hat\beta_{Xn},\hat\beta_{Yn})\}$, if the weights are equal for clusters. We also consider other weighting approaches in the estimation procedures for $\gamma_b$. Let $w_{i\cdot}$ denote the weight of cluster $i$ and $w_{i\cdot} =  \sum_{j=1}^{k_i} w_{ij}$. A nonparametric estimator of the CDF of $\beta_X$ is $\hat F_{\beta_X}(t) = \sum_{i=1}^n w_{i\cdot}I(\hat \beta_{Xi} \leq t)$, similarly $\hat F_{\beta_X}(t-) = \sum_{i=1}^n w_{i\cdot}I(\hat \beta_{Xi} < t)$, and we define $\hat F_{\beta_X}^*(t) = \big\{\hat F_{\beta_X}(t)+\hat F_{\beta_X}(t-)\big\}/2$. Estimation for $F_{\beta_Y}^*$ is similar. Therefore, one estimator of $\gamma_b$ is 
$$\hat{\gamma}_{b_M} = \frac{\sum\limits_{i=1}^n w_{i\cdot} \Big[\hat F^*_{\beta_{X}}(\hat \beta_{Xi}) -  \bar{\hat{F}}^*_{\beta_{X}}\Big]\Big[\hat F^*_{\beta_{Y}}(\hat \beta_{Yi}) -  \bar{\hat{F}}^*_{\beta_{Y}}\Big]}{\sqrt{\sum\limits_{i=1}^n w_{i\cdot} \Big[\hat F^*_{\beta_{X}}(\hat \beta_{Xi}) -  \bar{\hat{F}}^*_{\beta_{X}}\Big]^2}\sqrt{\sum\limits_{i=1}^n w_{i\cdot} \Big[\hat F^*_{\beta_{Y}}(\hat \beta_{Yi}) -  \bar{\hat{F}}^*_{\beta_{Y}}\Big]^2}},$$
where $\hat{\beta}_{X1}=\hat{\beta}_{Y1}=0$, $\bar{\hat{F}}^*_{\beta_{X}} = \sum\limits_{i=1}^n  w_{i\cdot} \hat F^*_{\beta_{X}}(\hat \beta_{Xi})$, and $\bar{\hat{F}}^*_{\beta_{Y}} = \sum\limits_{i=1}^n  w_{i\cdot} \hat F^*_{\beta_{Y}}(\hat \beta_{Yi})$. When $w_{i\cdot} = 1/n$, $\bar{\hat{F}}^*_{\beta_{X}} = \bar{\hat{F}}^*_{\beta_{Y}} = 1/2$. 

If cluster sizes are very small, the estimates of $\beta_X$ and $\beta_Y$, and thus $\hat{\gamma}_{b_M}$, may be poor. We consider another estimation approach. As shown in Section \ref{relationship} equation (\ref{spearmanrelation}), $\gamma_t$ is approximated by a linear combination of $\gamma_w$ and $\gamma_b$, where the weights are functions of the rank ICCs, $\gamma_{Ix}$ and $\gamma_{Iy}$. We can use this relationship to obtain an estimate of $\gamma_b$, 
$$\hat{\gamma}_{b_A} = \frac{\hat \gamma_{t} -\hat \gamma_{w}\sqrt{(1-\hat \gamma_{I_X} + \hat D_{X})(1-\hat \gamma_{I_Y} +\hat D_{Y})}}{\sqrt{(\hat \gamma_{I_X}-\hat D_{X})(\hat \gamma_{I_Y}-\hat D_{Y})}},$$
where $\hat \gamma_{I_X}$ and $\hat \gamma_{I_Y}$ are nonparametric estimators of $\gamma_{I_X}$ and $\gamma_{I_Y}$ \citep{tu2023}, $\hat D_{X} = \frac{\sum_{i=1}^n w_{i\cdot} \sum_{j<j'} \frac{2}{k_i(k_i-1)} \big[\hat F^*(x_{ij}) - \bar F^*_i\big]\big[\hat F^*(x_{ij'}) - \bar F^*_i\big]}{\sum_{i=1}^n \sum_{j=1}^{k_i} w_{ij} \big[\hat F^*(x_{ij}) - \bar F^*\big]^2}$, $\bar F^*_i = \sum_{j}F^*(x_{ij})/k_i$, and similar for $\hat D_{Y}$. If the cluster size in the population is infinite, $D_{X} = D_{Y} = 0$, then $\hat \gamma_{b_A} = \frac{\hat \gamma_{t} -\sqrt{(1-\hat \gamma_{Ix})(1-\hat \gamma_{Iy})}\hat \gamma_{w}}{\sqrt{\hat \gamma_{Ix} \hat \gamma_{Iy}}}$. Note that $\hat \gamma_{b_A}$ can be greater than 1 or less than $-1$; in those cases, we define $\hat \gamma_{b_A}$ to be $1$ or $-1$, respectively. As shown in the simulations, when cluster sizes are very small, $\hat{\gamma}_{b_A}$ may be preferable over $\hat{\gamma}_{b_M}$. If either of the rank ICCs is very small, $\sqrt{\hat \gamma_{Ix} \hat \gamma_{Iy}} \approx 0$ and $\hat \gamma_{b_A}$ can be unstable.

\section{Inference}
\label{inference}
The large sample distribution of $\hat \gamma_w$ can be obtained by bootstrapping or large sample approximation. Here we focus on the large sample approach using M-estimation \citep{stefanski2002}. The CPM is fit by minimizing the multinomial/nonparametric likelihood, and then the variance of parameter estimates can be estimated using a sandwich variance estimator that accounts for clustering. This is equivalent to fitting generalized estimating equation (GEE) methods for ordinal response variables with independence working correlation \citep{tian2023}. Let $\psi_X(\cdot) = \mathbf{U}_X(\boldsymbol{\theta})$ denote
the estimating function for the CPM of $X$ on $Z$ with a vector of parameters $\boldsymbol{\theta}_X$, and $\psi_Y(\cdot) = \mathbf{U}_Y(\boldsymbol{\theta})$ denote the estimating function for the CPM of $Y$ on $Z$ with a vector of parameters $\boldsymbol{\theta}_Y$. See the Supporting Information for details about these estimating functions. The components necessary for computing $\gamma_w$ are denoted by $\theta_{w1}$, $\theta_{w2}$, $\theta_{w3}$, $\theta_{w4}$, and $\theta_{w5}$ such that $\gamma_w = (\theta_{w3} -\theta_{w1}\theta_{w2})/\sqrt{(\theta_{w4}-\theta_{w1}^2)(\theta_{w5} - \theta_{w2}^2)}$, where $\theta_{w1} = E(X_{ij,res})$, $\theta_{w2} = E(Y_{ij,res})$, $\theta_{w3} = E(X_{ij,res}Y_{ij,res})$, $\theta_{w4} = E(X_{ij, res}^2)$, $\theta_{w5} = E(Y_{ij, res}^2)$. We can stack
$\psi_X(\cdot)$ and $\psi_Y(\cdot)$ together with these components and then have the following estimating function, 
\begin{align*}
& \psi_w(\mathbf{X}_i, \mathbf{Y}_i, \mathbf{Z}_i, \boldsymbol{\theta}_w) \\
& = \Big\{\psi_X(\mathbf{X}_i, \mathbf{Z}_i, \boldsymbol{\theta}_X), \psi_Y(\mathbf{Y}_i, \mathbf{Z}_i, \boldsymbol{\theta}_Y), 
I_i^T\mathbf{X}_{i,res}/k_i - \theta_{w1},  I_i^T\mathbf{Y}_{i,res}/k_i - \theta_{w2}, \mathbf{X}_{i,res}^T\mathbf{Y}_{i,res}/k_i - \theta_{w3}, \\ 
& \:\:\:\:\:\: \mathbf{X}_{i,res}^T\mathbf{X}_{i,res}/k_i  - \theta_{w4},\mathbf{Y}_{i,res}^T\mathbf{Y}_{i,res}/k_i - \theta_{w5}\Big\}^T,
\end{align*}
where $\boldsymbol{\theta}_w = (\boldsymbol{\theta}_X, \boldsymbol{\theta}_Y, \theta_{w1}, \theta_{w2}, \theta_{w3}, \theta_{w4} ,\theta_{w5})$, $I_i$ is a vector of ones with a length of $k_i$. The estimating equations are $\sum_{i=1}^n \psi_w(\mathbf{X}_i, \mathbf{Y}_i, \mathbf{Z}_i; \hat{\boldsymbol{\theta}}_w) = 0$. Under standard regularity conditions \citep{stefanski2002}, then we have 
$$\sqrt{n}(\hat{\boldsymbol{\theta}}_w-\boldsymbol{\theta}_w )  \stackrel{d}{\rightarrow} MVN\big(\mathbf{0}, \mathbf{V}(\boldsymbol{\theta}_w)\big),$$ where $\mathbf{V}(\boldsymbol{\theta}_w) =  \mathbf{A}(\boldsymbol{\theta}_w)^{-1} \mathbf{B}(\boldsymbol{\theta}_w)\big\{\mathbf{A}(\boldsymbol{\theta}_w)^{-1}\big\}^{T}$, $\mathbf{A}(\boldsymbol{\theta}_w) = E\big[- \partial \psi_{w}(\mathbf{X}_i, \mathbf{Y}_i, \mathbf{Z}_i,\boldsymbol{\theta}_w)/\partial \boldsymbol{\theta}_w^T\big]$, and $\mathbf{B}(\boldsymbol{\theta}_w) = E\big[\psi_w(\mathbf{X}_i, \mathbf{Y}_i, \mathbf{Z}_i,\boldsymbol{\theta}_w)\psi_w(\mathbf{X}_i, \mathbf{Y}_i, \mathbf{Z}_i,\boldsymbol{\theta}_w)^T\big]$. Since our estimator of $\gamma_w$ is a function of $\hat \theta_{w1}$, $\hat \theta_{w2}$, $\hat \theta_{w3}$, $\hat \theta_{w4}$, and $\hat \theta_{w5}$, the delta method can be used to obtain its large sample distribution. Then we can compute the asymptotic standard error (SE) of $\gamma_w$ and construct confidence intervals (CIs) for $\gamma_w$. We also can easily show that $\hat \gamma_w$ is a consistent estimator of $\gamma_w$. 

We use a similar approach to obtain the large sample distribution of $\hat \gamma_{b_M}$.  Let
$\boldsymbol{\beta}_X = (0, \beta_{X2},..., \beta_{Xn})^T$ and $\boldsymbol{\beta}_Y=(0, \beta_{Y2},..., \beta_{Yn})^T$ denote the coefficients of cluster index in the CPMs of $X$ and $Y$, respectively. Note that the coefficient of the reference cluster is zero. To obtain the asymptotic variance of $\gamma_b$, we treat $\boldsymbol{\beta}_X$ and $\boldsymbol{\beta}_Y$ as random effects, for simplicity assuming that $\beta_{Xi} \stackrel{i.i.d}{\sim} N(\mu_{\beta_X}, \sigma^2_{\beta_X})$ and $\beta_{Yi} \stackrel{i.i.d}{\sim} N(\mu_{\beta_Y}, \sigma^2_{\beta_Y})$.  The components necessary for computing $\gamma_b$ are denoted by $\theta_{b1}$, $\theta_{b2}$, $\theta_{b3}$, $\theta_{b4}$, and $\theta_{b5}$ such that $\gamma_b = (\theta_{b3} -\theta_{b1}\theta_{b2})/\sqrt{(\theta_{b4}-\theta_{b1}^2)(\theta_{b5} - \theta_{b2}^2)}$, where $\theta_{b1} = E\big[F_{\beta_{X}}(\beta_{Xi})\big]$,
$\theta_{b2} = E\big[F_{\beta_{Y}}(\beta_{Yi})\big]$,
\\ $\theta_{b3}= E\big[F_{\beta_{X}}(\beta_{Xi}) F_{\beta_{Y}}(\beta_{Yi})\big]$, $\theta_{b4} = E\Big\{\big[F_{\beta_{X}}(\beta_{Xi})\big]^2\Big\}$, $\theta_{b5} = E\Big\{\big[F_{\beta_{Y}}^2(\beta_{Yi})\big]^2\Big\}$, and $F_{\beta_{X}}$ and $F_{\beta_{Y}}$ are the CDFs of normal distributions. Note that $E\big[F_{\beta_{X}}(\beta_{Xi})\big] = E\big[F_{\beta_{X}}(\beta_{Yi})\big]=1/2$ in theory but they may not be $1/2$ in estimation if $w_{i\cdot} \neq 1/n$. Similar to the inference procedure of $\gamma_w$ above, we stack
$\psi_X(\cdot)$ and $\psi_Y(\cdot)$ with the components needed to compute $\gamma_b$ stacked together, yielding the following estimating function, 
\begin{align*}
& \psi_b(\mathbf{X}_i, \mathbf{Y}_i, \mathbf{Z}_i, \boldsymbol{\theta}_b) \\
& = \Big\{\psi_X(\mathbf{X}_i, \mathbf{Z}_i, \boldsymbol{\theta}_X), \psi_Y(\mathbf{Y}_i, \mathbf{Z}_i, \boldsymbol{\theta}_Y), \beta_{Xi} - \mu_{\beta_X}, \beta_{Yi} - \mu_{\beta_Y}, \beta_{Xi}^2 - M_{\beta_X} , \beta_{Yi}^2 -  M_{\beta_Y}, \\
& F_{\beta_X}(\beta_{Xi}) - \theta_{b1}, F_{\beta_Y}(\beta_{Yi})- \theta_{b2}, F_{\beta_X}(\beta_{Xi}) F_{\beta_Y}(\beta_{Yi}) - \theta_{b3}, \big[F_{\beta_X}(\beta_{Xi})\big]^2 - \theta_{b4}, \big[F_{\beta_X}(\beta_{Yi})\big]^2- \theta_{b5}\Big\}^T,\end{align*}
where $\boldsymbol{\theta}_b = (\boldsymbol{\theta}_X, \boldsymbol{\theta}_Y, \mu_{\beta_X}, \mu_{\beta_Y},M_{\beta_X}, M_{\beta_Y}, \theta_{b1}, \theta_{b2}, \theta_{b3}, \theta_{b4} ,\theta_{b5})$,
$M_{\beta_X} = E( \beta_{X}^2) = \mu_{\beta_X}^2+ \sigma^2_{\beta_X}$, and $M_{\beta_Y} = E( \beta_{Y}^2) = \mu_{\beta_Y}^2 + \sigma^2_{\beta_Y}$. The estimating equations are $\sum_{i=1}^n \psi_b(\mathbf{X}_i, \mathbf{Y}_i, \mathbf{Z}_i; \hat{\boldsymbol{\theta}}_b) = 0$. We have 
$$\sqrt{n}(\hat{\boldsymbol{\theta}}_b -\boldsymbol{\theta}_b)  \stackrel{d}{\rightarrow} MVN\big(\mathbf{0}, \mathbf{V}(\boldsymbol{\theta}_b)\big)$$ under standard regularity conditions \citep{stefanski2002}, where \\ $\mathbf{V}(\boldsymbol{\theta}_b) =  \mathbf{A}(\boldsymbol{\theta}_b)^{-1} \mathbf{B}(\boldsymbol{\theta}_b)\big\{\mathbf{A}(\boldsymbol{\theta}_b)^{-1}\big\}^{T}$, $\mathbf{A}(\boldsymbol{\theta}_b) = E\big[- \partial \psi_{b}(\mathbf{X}_i, \mathbf{Y}_i, \mathbf{Z}_i, \boldsymbol{\theta})/\partial \boldsymbol{\theta}^T_b\big]$, and $\mathbf{B}(\boldsymbol{\theta}) = E\big[\psi_b(\mathbf{X}_i, \mathbf{Y}_i, \mathbf{Z}_i,\boldsymbol{\theta}_{b})\psi_b(\mathbf{X}_i, \mathbf{Y}_i, \mathbf{Z}_i,\boldsymbol{\theta}_{b})^T\big]$. The large sample distribution of $\hat \gamma_{b_M}$ can be derived from the large sample distribution of $\hat{\boldsymbol{\theta}}_b$ using the delta method. Then it is easy to show that $\hat \gamma_{b_M}$ is a consistent estimator of $\gamma_{b_M}$. We also use this expression for inference for $\hat \gamma_{b_A}$.

As mentioned in Section \ref{estimation}, if the same weight is assigned to all the observations, our estimator of the total Spearman rank correlation $\hat \gamma_t$ equals the estimator of Shih and Fay \citeyearpar{shih2017}. Hence, we adapt the inference method of Shih and Fay \citeyearpar{shih2017} via incorporating weighting into the estimation procedures to obtain the asymptotic variance of $\hat \gamma_t$. Shih and Fay \citeyearpar{shih2017} have provided an analytical form for the asymptotic distribution of the estimator of $\gamma_t$, which is a function of $X_{ij}$, $Y_{ij}$, $F_X$, $F_Y$, and $F_{XY}$. The asymptotic variance can be estimated with $\hat F_X$, $\hat F_Y$, and $\hat F_{XY}$ plugged in for $F_X$, $F_Y$, and $F_{XY}$. Here we allow $\hat F_X$, $\hat F_Y$, and $\hat F_{XY}$ to be obtained based on either assigning equal weights to observations or assigning equal weights to clusters, and then plug them in to estimate the asymptotic variance of $\hat \gamma_t$.

The confidence intervals constructed by our variance estimators might not always fall within the natural range, $[-1, 1]$, particularly when the rank correlation estimate is near the boundaries (i.e., -1 or 1). To address this, one can apply the Fisher transformation, $\frac{1}{2}\text{log}\frac{1+r}{1-r}$. Specifically, first obtain the rank correlation estimate and its variance on the original scale, then compute the estimate and the variance (via the delta method) on the transformed scale, calculate the confidence interval on this scale, and finally transform the bounds of this confidence interval back to the original scale. In our R package \texttt{rankCorr} \citep{tuRankCorr}, the function for computing the rank correlations, \texttt{rankCorrCluster()}, allows users to choose whether to apply Fisher transformation to compute confidence intervals. When the number of clusters is small, to account for additional uncertainty, we can use Student's t-distribution with our variance estimators to construct confidence intervals. 

\section{Simulations}
\label{simulations}
\subsection{Continuous data}
\label{simcont}
We used a bivariate additive model for generating latent variables: $\begin{pmatrix} X_{0ij} \\ Y_{0ij} \end{pmatrix} = \begin{pmatrix} U_{Xi}\\ U_{Yi} \end{pmatrix} + \begin{pmatrix} R_{Xij} \\ R_{Yij}\end{pmatrix},$
where $\begin{pmatrix} U_{Xi} \\ U_{Yi} \end{pmatrix}  \stackrel{i.i.d}{\sim}  N \begin{pmatrix} \begin{pmatrix} 1 \\ -1 \end{pmatrix}, \begin{pmatrix} 1 & \rho_{0b} \\   \rho_{0b} & 1 \end{pmatrix} \end{pmatrix} $, $\begin{pmatrix} R_{Xij}\\ R_{Yij} \end{pmatrix} \stackrel{i.i.d}{\sim}  N  \begin{pmatrix} \begin{pmatrix} 0 \\ 0 \end{pmatrix}, \begin{pmatrix} 1 & \rho_{0w} \\   \rho_{0w} & 1 \end{pmatrix} \end{pmatrix}$. Here, $\rho_{0t} = (\rho_{0b} + \rho_{0w})/2$. Let $(X_{ij}, Y_{ij})$ be the observation of the $j$th individual in the $i$th cluster, where $i=1,2,....,n$; $j=1,2,...,k_i$; and $k_i$ is the size of the $i$th cluster. We considered three scenarios: (I) $X_{ij}=X_{0ij}$ and $Y_{ij}=Y_{0ij}$; (II) $X_{ij}=X_{0ij}$ and $Y_{ij}=\exp(Y_{0ij})$; (III) $X_{ij}=\exp(U_{Xi})+R_{Xij}$ and $Y_{ij}=\exp(\exp(U_{Yi})+R_{Yij})$. Under Scenarios I and II, since $(X_{0ij}, Y_{0ij})^T$ is bivariate normal, the true total, between-, and within-cluster Spearman rank correlations are $\gamma_t = 6\arcsin(\rho_{0t}/2)/\pi$, $\gamma_b = 6\arcsin(\rho_{0b}/2)/\pi$, and $\gamma_w = 6\arcsin(\rho_{0w}/2)/\pi$ \citep{Pearson1907}. Under Scenario III,  $\gamma_b$ and $\gamma_w$ are the same as those in Scenarios I and II, but $\gamma_t$ is different because $(X_{ij}, Y_{ij})^T$ is not normally distributed. We empirically computed $\gamma_{t}$ under Scenario III by generating one million clusters each with 100 observations, and then computing $\gamma_t$. We also empirically computed the total, between- and within-cluster Pearson correlations (i.e., $\rho_t$, $\rho_b$, and $\rho_w$) under Scenarios II and III. While $\gamma_b$ and $\gamma_w$ are identical under the three scenarios, $\rho_b$ and $\rho_w$ are sensitive to skewness and depend on the scale of interest (Table \ref{tab1}). In Scenarios I and II, the rank ICCs of $X$ and $Y$ are both 0.48. In Scenario III, the rank ICC of $X$ is 0.97 while that of $Y$ is 0.37. Our simulations considered various configurations of sample cluster size under Scenarios I and II: $k_i=10$, $k_i=20$, $k_i=30$, and $k_i$ uniformly ranging from 1 to 50. In Scenario III, we set $k_i=20$. The simulations were replicated 1000 times for each configuration under each scenario at $n=100$, and $(\rho_{0b},\rho_{0w}) \in \{(0.8, 0.7), (0.8, 0), (0, 0.8), (0, 0), (0.8, -0.7)\}$.

\begin{sidewaystable}
\small
\centering
\caption{The total, between-cluster, and within-cluster Spearman rank correlations ($\gamma_{t}, \gamma_{b}, \gamma_{w}$) and Pearson correlations ($\rho_{t}, \rho_{b}, \rho_{w}$) under Scenarios I, II, and III with 5 simulation settings}
\begin{tabular}[t]{@{\extracolsep{6pt}}>{\RaggedRight}p{3cm}>{\RaggedRight}p{3cm}>{\RaggedRight}p{3cm}>{\RaggedRight}p{3cm}>{\RaggedRight}p{3cm}>{\RaggedRight}p{3cm}}
\hline
\multicolumn{1}{c}{($\rho_{0t}, \rho_{0b}, \rho_{0w}$)} & \multicolumn{2}{c}{($\gamma_{t}, \gamma_{b}, \gamma_{w}$)} & \multicolumn{3}{c}{($\rho_{t}, \rho_{b}, \rho_{w}$)} \\
\cline{1-1} \cline{2-3} \cline{4-6}
\multicolumn{1}{c}{I, II, III} & \multicolumn{1}{c}{I, II} & \multicolumn{1}{c}{III} & \multicolumn{1}{c}{I}& \multicolumn{1}{c}{II}& \multicolumn{1}{c}{III}\\
\hline
(0.75, 0.80, 0.70) & (0.73, 0.79, 0.68) & (0.53, 0.79, 0.68) & (0.75, 0.80, 0.70) & (0.42, 0.61, 0.33) & (0.02, 0.03, 0)\\

(0.40, 0.80, 0) & (0.38, 0.79, 0) & (0.31, 0.79, 0) & (0.40, 0.80, 0) & (0.22, 0.06, 0) & (0.01, 0.02, 0)\\

(0.40, 0, 0.80) & (0.38, 0, 0.79) & (0.25, 0, 0.79) & (0.40, 0, 0.80) & (0.22, 0.01, 0.37) & (0, 0, 0)\\

(0, 0, 0) & (0, 0, 0) & (0, 0, 0) & (0, 0, 0) & (0, 0, 0) & (0, 0, 0)\\

(0.05, 0.80, -0.7) & (0.05, 0.79, -0.68) & (0.09, 0.79, -0.68) & (0.05, 0.80, -0.70) & (0.03, 0.59, -0.32) & (0.01, 0.01, 0)\\
\hline
\end{tabular}
\label{tab1}
\end{sidewaystable}

In the simulations, we compared our estimators with naive nonparametric estimators: $\hat{\gamma}_{b_n}$ estimated by Spearman rank correlation between sample cluster medians and $\hat{\gamma}_{w_n}$ estimated by the rank correlation of within-cluster deviations (differences) from sample cluster medians. In addition, we compared the Spearman rank correlations with the Pearson correlations in Scenario I. The estimators of $\rho_b$ and $\rho_w$ are based on one-way random effects models: $\rho_b$ is estimated by Pearson correlation between the estimated cluster means from the random effects models and $\rho_w$ is estimated by Pearson correlation between the individual deviations from the estimated cluster means \citep{snijders1999}.

\begin{figure}[h!]
     \centering
     \includegraphics[width=1\textwidth]{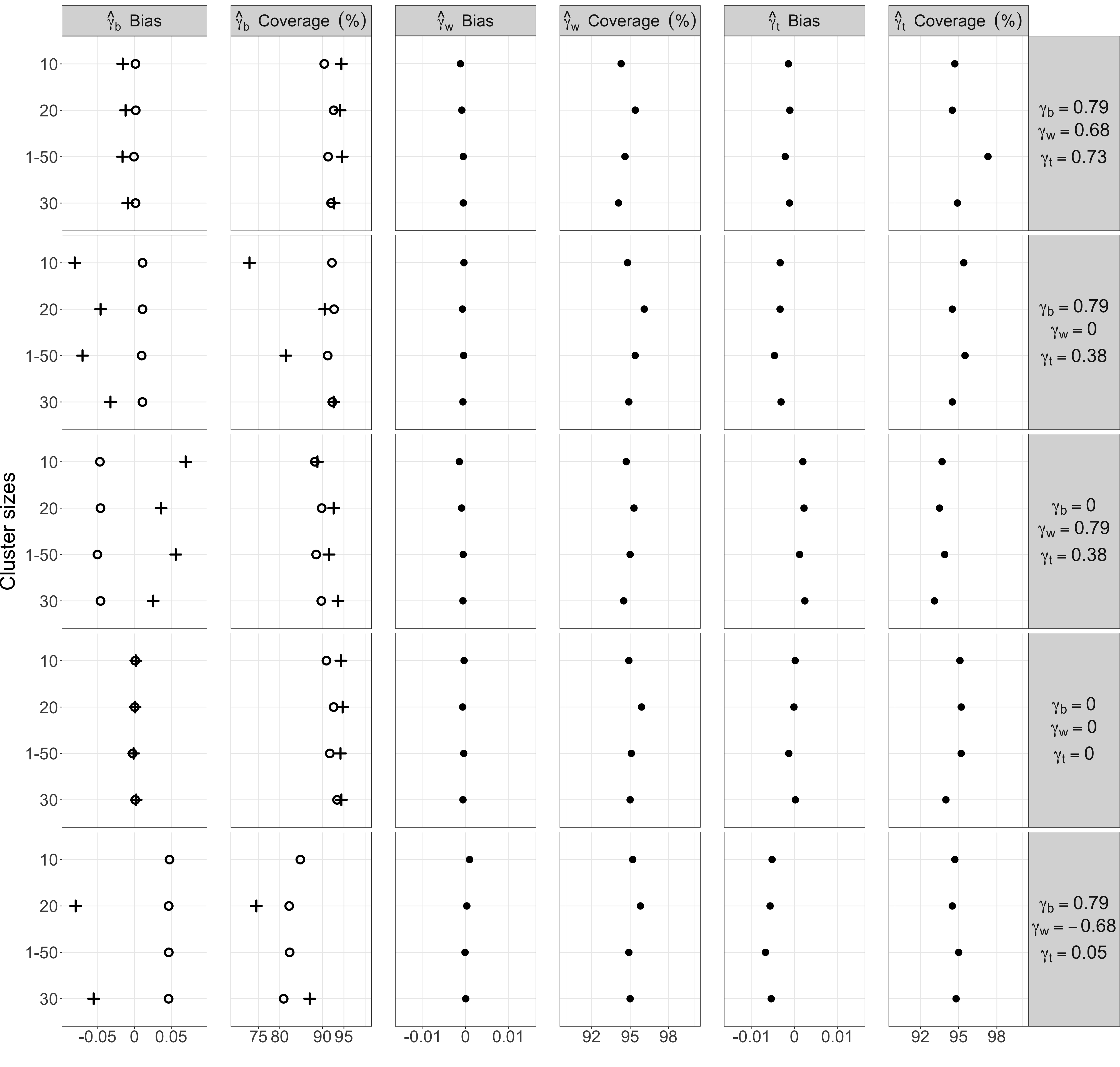}
     \caption[]{Bias and coverage of 95\% confidence intervals for our estimators of $\gamma_{b}$, $\gamma_{w}$, and $\gamma_{t}$ at different true values and different cluster sizes under Scenarios I (normality) and II (exponentiated $Y$). The circle sign stands for the approximation-based estimator ($\hat{\gamma}_{b_A}$) of $\gamma_b$, and the plus sign stands for the cluster-median-based estimator ($\hat{\gamma}_{b_M}$) of $\gamma_b$. The number of clusters was set at 100. ``1-50'' means the cluster size follows a uniform distribution from 1 to 50. When $(\gamma_b,\gamma_w, \gamma_t)=(0.79, -0.68, 0.05)$ and cluster sizes are 10 or 1-50, $\hat{\gamma}_{b_M}$ had negative bias greater than 0.05 and poor coverage below 75\% (not shown in the figure due to out of bounds). For specific bias and coverage values of $\hat{\gamma}_{b_A}$ and $\hat{\gamma}_{b_M}$, please refer to Web Table 1.}
        \label{fig:simcls}
\end{figure}

In summary, our estimators of $\gamma_b$, $\gamma_w$, and $\gamma_t$ had low bias and good coverage with modest numbers of clusters in Scenarios I and II (Figure \ref{fig:simcls}). They were also robust to the skewed data in Scenario II and they had lower bias than $\hat{\gamma}_{b_n}$ and $\hat{\gamma}_{w_n}$ (Web Table 2). In the extreme case where $\gamma_b$ and $\gamma_w$ are both strong but opposite (i.e., last row of Figure 2), our estimators of $\gamma_b$ were biased. In the other settings where $\gamma_b$ and $\gamma_w$ greatly differed (i.e., rows 2-3 of Figure 2), the estimators of $\gamma_b$ were also biased, although to a lesser extent. In these settings, the bias of $\hat \gamma_{b_A}$ was relatively smaller than that of $\hat \gamma_{b_M}$, particularly with small cluster sizes. As the cluster size increased, the bias of $\hat \gamma_{b_M}$ decreased, whereas the bias of $\hat \gamma_{b_A}$ remained relatively stable (also seen in Web Table 3). It is worth noting that in the extreme case (last row of Figure 2), the estimator of the between-cluster Pearson correlation based on random effects models also had similar bias, even when the data were normally distributed (Web Table 4). 

In Scenario III, our estimator of $\gamma_w$ still had low bias and good coverage (Web Table 5). In our setup, with $E[U_{Yi}] = -1$, our estimators of $\gamma_b$ and $\gamma_t$ had more bias under Scenario III than under Scenarios I and II (Web Table 5a). This is because $U_{Yi}$ was exponentiated in Scenario III, which led to cluster means that had a much smaller variance than that of the within-cluster deviations. In this setting, the within-cluster deviation often dominated the value of $Y_{ij}$ creating data where it is difficult to see the effect of clustering over the within-cluster variance. Our estimator of $\gamma_t$ struggled in this setting, producing biased estimates of $\gamma_t$ and thus biased estimates of $\gamma_b$ based on the approximation (\ref{spearmanrelation2}). In addition, estimation of $\gamma_b$ using $\hat{\gamma}_{b_M}$ also was biased, as estimated cluster medians, even with fairly larger cluster sizes, often were far from their true rankings due to the large residual noise. When $E[U_{Yi}]$ was changed from -1 to 1, the cluster means had a larger variance than that of the within-cluster deviations, leading to much smaller bias in the estimates of $\gamma_t$ and $\gamma_b$ (Web Table 5b). 

We also evaluated the performance of our estimators with small numbers of clusters. Here we used Student's t-distribution with $n-1$ degrees of freedom to construct confidence intervals. We conducted simulations with small numbers of clusters and large cluster sizes under Scenario II: $k_i=100$, $n \in \{10, 20\}$, and $(\rho_{0b},\rho_{0w}) \in \{(0.8, 0.7), (0.8, 0), (0, 0.8), (0, 0), (0.8, -0.7)\}$. The simulation results (Web Table 6) show that $\hat{\gamma}_w$ still had very low bias and good coverage in all cases. $\hat{\gamma}_{b_A}$, $\hat{\gamma}_{b_M}$, and $\hat{\gamma}_{t}$ had some bias, particularly in the settings with substantially different $\gamma_b$ and $\gamma_w$. $\hat{\gamma}_{b_A}$ had relatively lower bias than $\hat{\gamma}_{b_M}$, except when $\gamma_b=0$. The three estimators had coverage slightly below the nominal 0.95 level, which could be due to challenges of accurately estimating variance with a small number of clusters. In addition, we conducted simulations under Scenario II with varied $n$ and $k_i$ to explore the smallest number of clusters our method could handle: $n \in \{2, 10, 15, 25, 50, 100\}$, $k_i \in \{2, 5, 10, 15, 20, 30\}$, and $(\rho_{0b},\rho_{0w}) = (0.6, 0.5)$. The simulation results (Web Table 7) show that both $\hat{\gamma}_w$ and $\hat{\gamma}_t$ had low bias in nearly all cases, with the exception being settings with extremely low numbers of clusters ($n$=2) and cluster sizes (i.e., $k_i=2$ or 5). Except in these extreme cases, $\hat{\gamma}_w$ had good coverage. $\hat{\gamma}_t$ required $\geq 10$ clusters to ensure good coverage. $\hat{\gamma}_{b_A}$ had low bias with $\geq 10$ clusters, whereas $\hat{\gamma}_{b_M}$ required $\geq 25$ clusters to achieve low bias. Both $\hat{\gamma}_{b_A}$ and $\hat{\gamma}_{b_M}$ required $\geq 25$ clusters to ensure good coverage. 

Additional simulations were conducted to evaluate the performance of our estimators when the rank ICC was negative, which occurs when $k_i=2$. Our estimators of $\gamma_t$, $\gamma_b$, and $\gamma_w$ had very low bias and good coverage. Details are in the Supporting Information Web Table 8.

Furthermore, we investigated the performance of our estimators when the link function of the CPM was misspecified as logit, loglog, and cloglog under Scenario II. We conducted 1000 simulations at $n=100$ and $k_i=20$. Our estimators of $\gamma_b$, $\gamma_w$, and $\gamma_t$ performed similarly under the logit link as they did under the correct probit link function (Web Table 9). When the link function was misspecified as loglog or cloglog, if $\gamma_w$ was large and had the opposite direction of $\gamma_b$, our estimator of $\gamma_w$ had bias toward the direction of $\gamma_b$. 

\subsection{Ordered categorical data}
\label{simordinal}
We also evaluated the performance of our estimators for ordered categorical data. We simulated 5-level and 10-level ordered categorical data by discretizing $X_{ij}$ and $Y_{ij}$ in Scenario I (described in Section \ref{simcont}) with cutoffs at quantiles (i.e., using the 0.2, 0.4, 0.6, 0.8 quantiles for 5 levels; and the 0.1, 0.2, ..., 0.8, 0.9 quantiles for 10 levels). We empirically computed $\gamma_t$, $\gamma_b$, and $\gamma_w$ by generating one million clusters and 100 observations per cluster, with cluster medians analytically derived. The values of $\gamma_t$, $\gamma_b$, and $\gamma_w$ of the 10-level ordered categorical variables are close to those of the continuous variables, while those of the 5-level ordered categorical variables are slightly smaller (Web Table 10). We conducted 1000 simulations at $n=100$ and $k_i=20$. Our estimators of $\gamma_b$, $\gamma_w$, and $\gamma_t$ had very low bias and good coverage (Web Table 10). When $\gamma_b$ was large, $\hat{\gamma}_{b_A}$ had bias, which might be due to equation (\ref{spearmanrelation2}) being a poor approximation of $\gamma_t$ with ordered categorical data. This bias decreased as the number of ordered categories increased.

\section{Applications}
\subsection{Longitudinal biomarker data}
\begin{figure}[h!]
     \centering
     \begin{subfigure}[b]{0.333\textwidth}
         \centering 
         \includegraphics[width=1\textwidth]{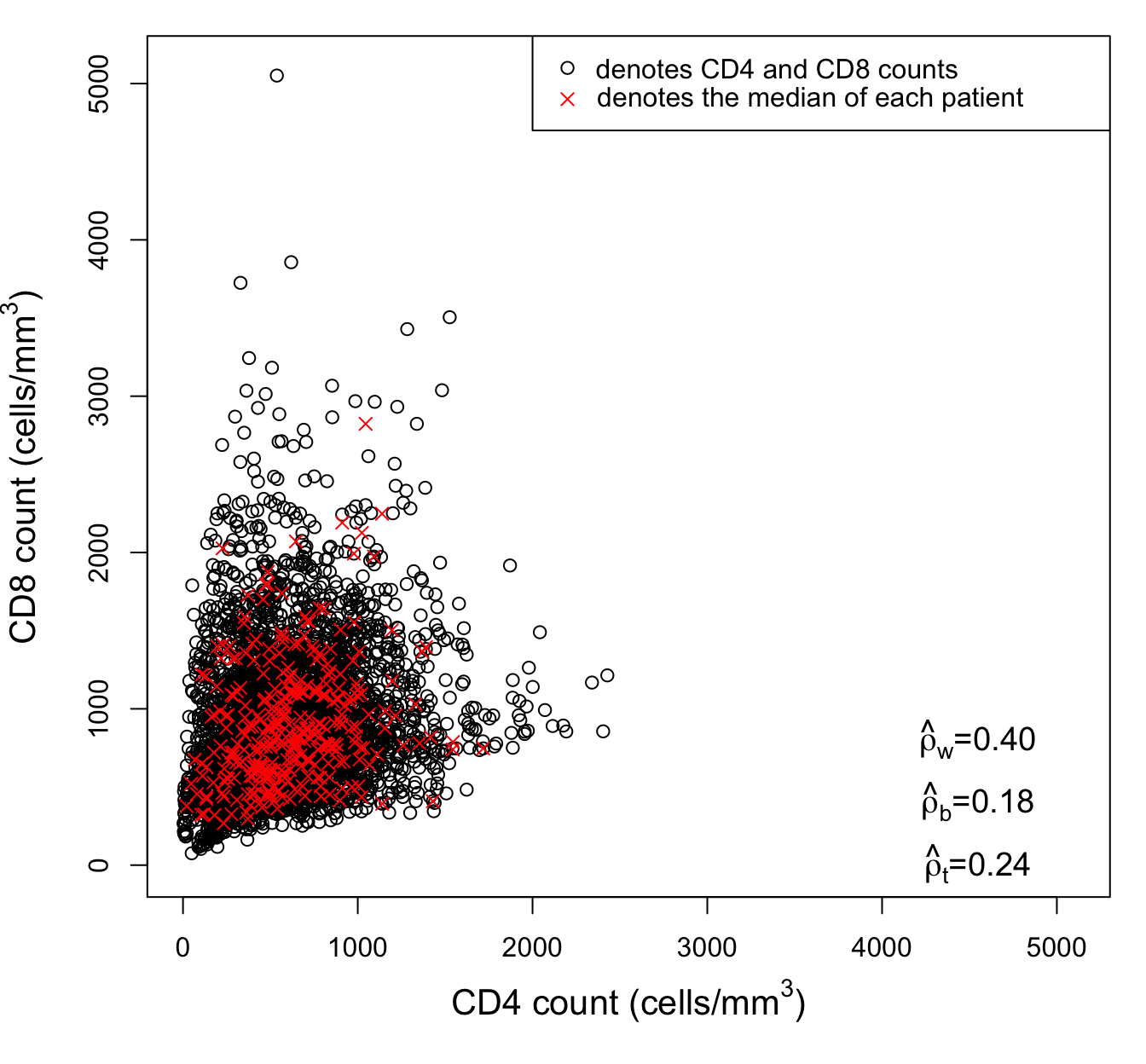} 
         \caption{Original scale} 
         \label{fig:cd4} 
     \end{subfigure}\hfill
          \begin{subfigure}[b]{0.333\textwidth}
         \centering
         \includegraphics[width=1\textwidth]{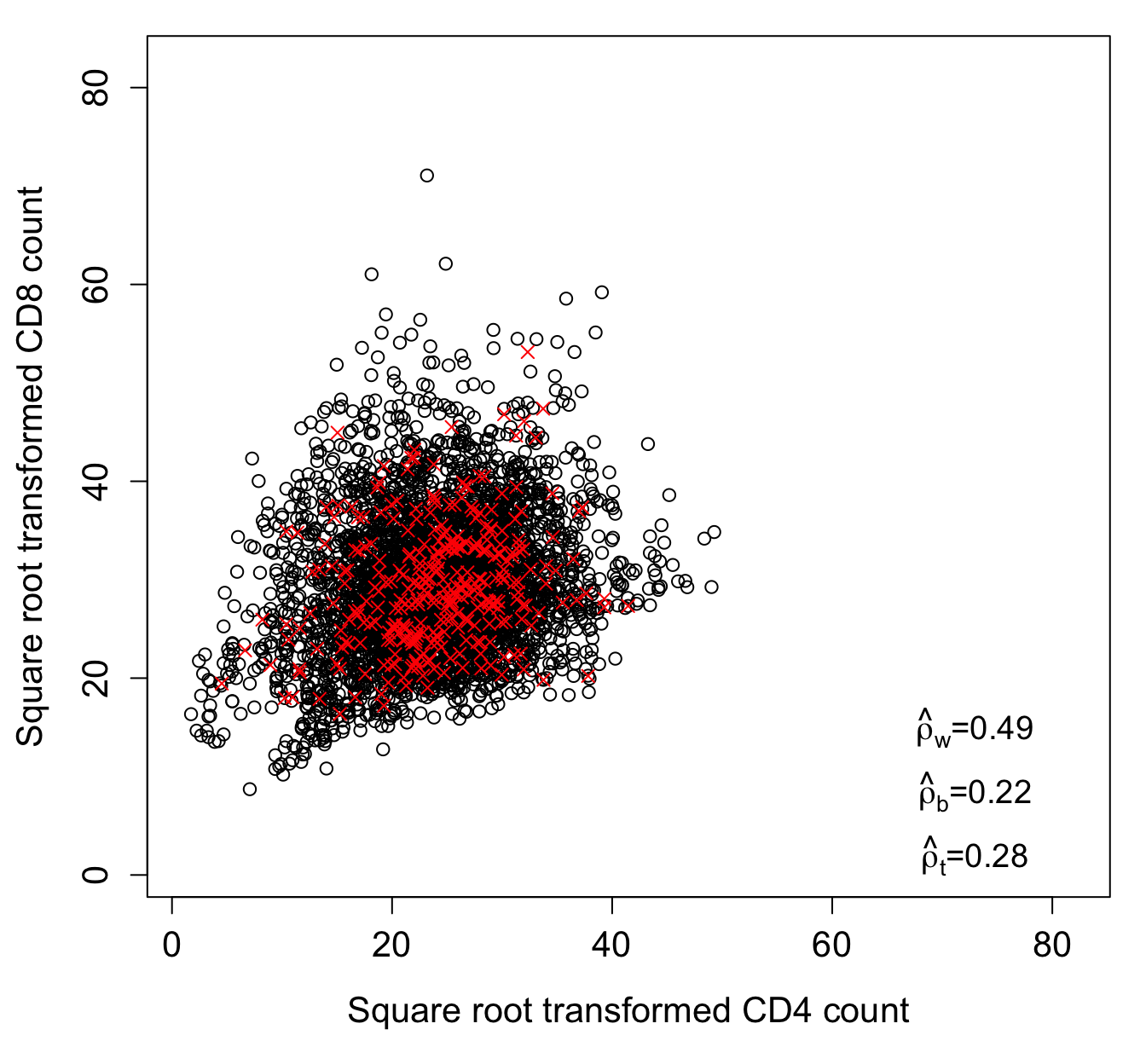} 
         \caption{Square-root transformation}
         \label{fig:cd4sqrt} 
     \end{subfigure}\hfill
    \begin{subfigure}[b]{0.333\textwidth}
         \centering
         \includegraphics[width=1\textwidth]{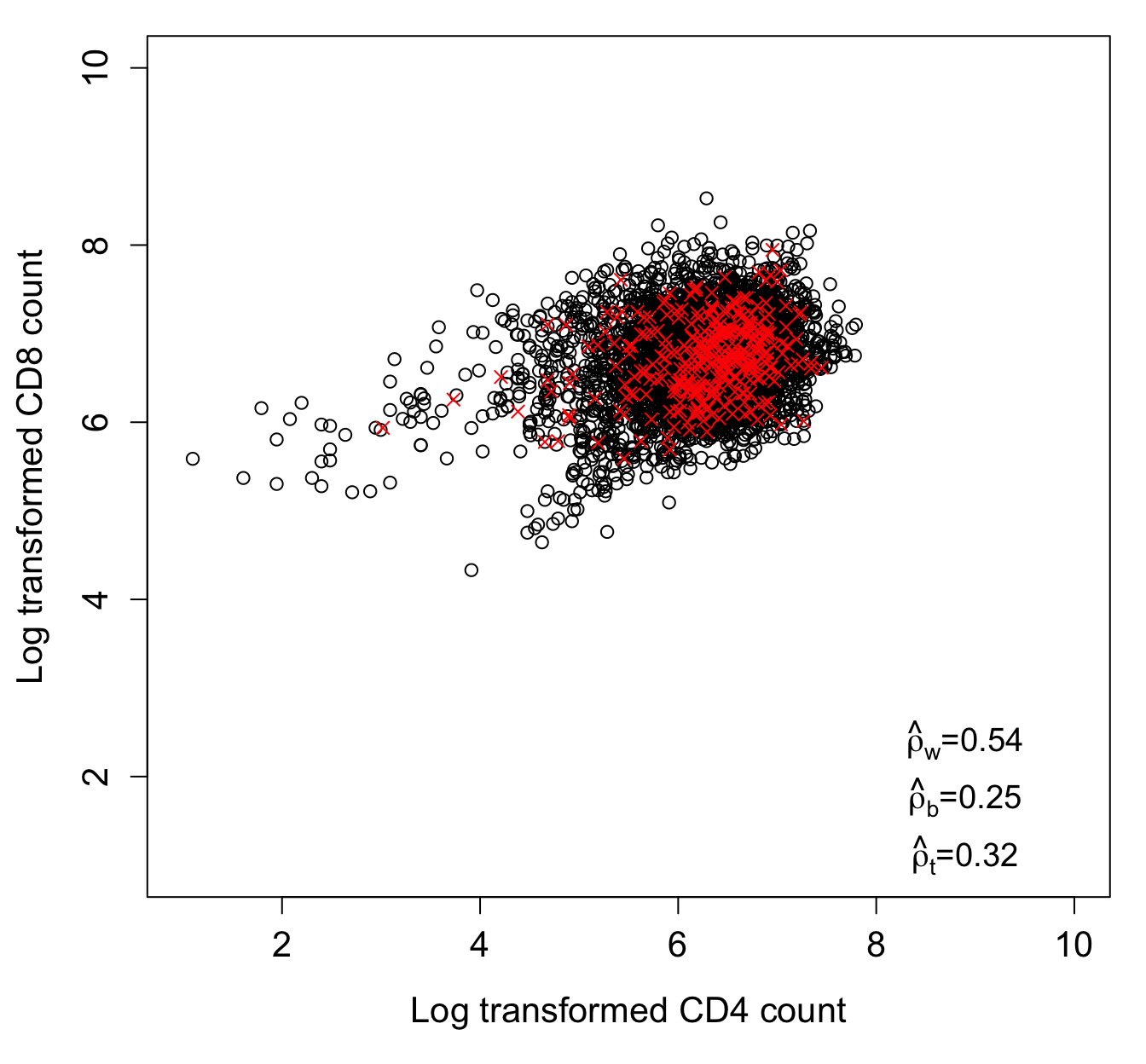} 
         \caption{Log transformation}
         \label{fig:cd4log}
     \end{subfigure}
     \caption[]{Scatter plot of CD4 and CD8 counts (cells/mm$^3$) and estimates of within-, between-cluster, and total Pearson ($\rho_w$, $\rho_b$, $\rho_t$) correlations. The red cross sign represents the sample cluster median and the circle sign represents the observation. The estimates (95\% confidence intervals) of within-, between-cluster, and total Spearman rank correlations are invariant to data transformations, $\hat{\gamma}_w = 0.53\:[0.51,0.55]$, $\hat{\gamma}_{b_M}=0.24 \:[0.20,0.29]$, $\hat{\gamma}_{b_A}=0.21 \:[0.17,0.26]$, and $\hat{\gamma}_t = 0.29 \:[0.25,0.32]$, where $\hat{\gamma}_{b_{M}}$ is the estimator of $\gamma_b$ based on cluster medians obtained from CPM and $\hat{\gamma}_{b_A}$ is the estimator of $\gamma_b$ based on the linear approximation.}
        \label{fig:example1}
\end{figure}

Repeated measures of CD4 and CD8 lymphocyte counts (cells/mm$^3$) were taken on 325 women living with HIV who started antiretroviral therapy (ART) at the Vanderbilt Comprehensive Care Clinic between 1998 and 2012 \citep{castilho2016}. There is interest in evaluating the correlation between same-day CD4 and CD8 counts while considering the potential clustering in the data. All same-day CD4 and CD8 measurements taken within $\pm 6$ months of the baseline date were included in analyses; the number of observations per woman ranged from 1 to 54. In this case, the cluster is the person, so it makes sense to assign equal weights to people rather than measurements. The data were very skewed, especially the CD8 count (Figure \ref{fig:example1}). 

The rank ICC estimates of CD4 and CD8 counts were 0.77 and 0.76, respectively, suggesting strong similarity between measurements from the same woman. The within-cluster Spearman rank correlation estimate was 0.53 (95\% CI: [0.51,0.55]), suggesting moderate correlation between the fluctuations in the repeated CD4 and CD8 measurements. The between-cluster Spearman rank correlation was estimated to be $\hat{\gamma}_{b_M}=0.24$ (95\% CI: [0.20,0.29]) via cluster medians obtained from CPMs and was estimated to be $\hat{\gamma}_{b_A}=0.21$ (95\% CI: [0.17,0.26]) via the approximation approach, indicating a weak but positive correlation between median CD4 and CD8 counts. The Spearman rank correlation between the sample cluster medians was 0.24, close to our between-cluster Spearman rank correlation estimates. The total Spearman rank correlation estimate, 0.29 (95\% CI: [0.25,0.32]), suggests a weak to moderate overall correlation after combining between-cluster and within-cluster correlations.

In contrast, the Pearson correlations are highly sensitive to extreme values and vary with data transformation. The within-cluster, between-cluster, and total Pearson correlation estimates on the original scale obtained from a random effects model were estimated to be 0.40, 0.18, and 0.24, respectively, which were impacted by some extreme measurements. The three Pearson correlations were estimated to be 0.49, 0.22, and 0.28, respectively, after square root transformation, and 0.54, 0.25, and 0.32, respectively, after log transformation. These notable differences in the three Pearson correlation estimates after data transformation (e.g., $\hat{\rho}_w$ varying from 0.40 to 0.54) demonstrate the sensitivity of Pearson correlation to the choice of scale. It is uncertain whether square root or log transformations are the most appropriate, or whether other transformations should be considered. It is also unclear whether the same transformation should be applied to both variables or whether each variable requires a different transformation. In contrast, our estimates of within-cluster, between-cluster, and total Spearman rank correlations are invariant to any choice of scale. Our estimators offer a robust framework where one does not need to consider transformations. 

\subsection{Cluster randomized controlled trial data}
\begin{figure}
     \centering
     \begin{subfigure}[b]{0.5\textwidth}
         \centering
         \includegraphics[width=0.9\textwidth]{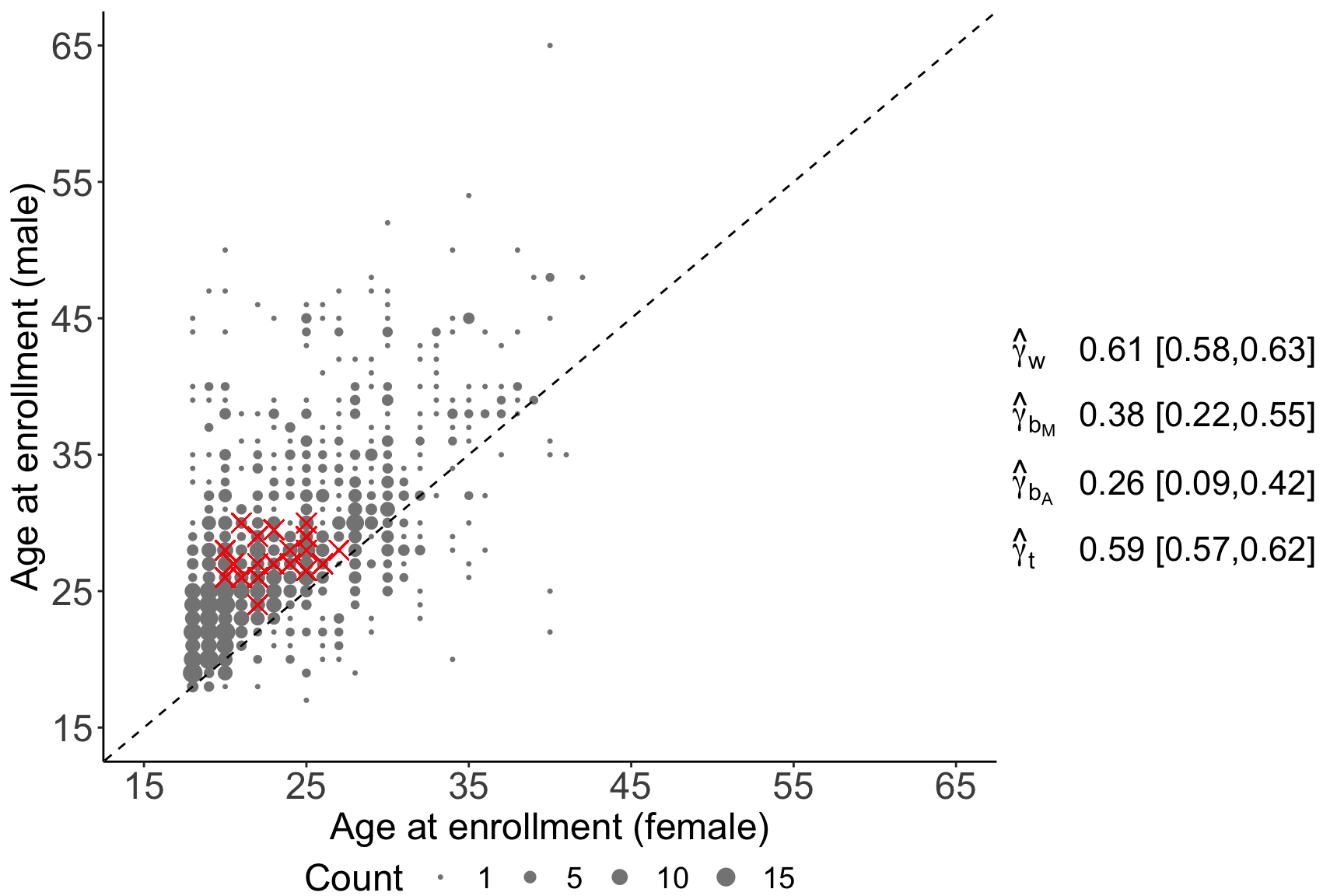}\vspace{-0.1cm}
         \caption{Age at enrollment}\vspace{0.3cm}
         \label{fig:age}
     \end{subfigure}\hfill
     \begin{subfigure}[b]{0.5\textwidth}
         \centering
         \includegraphics[width=0.9\textwidth]{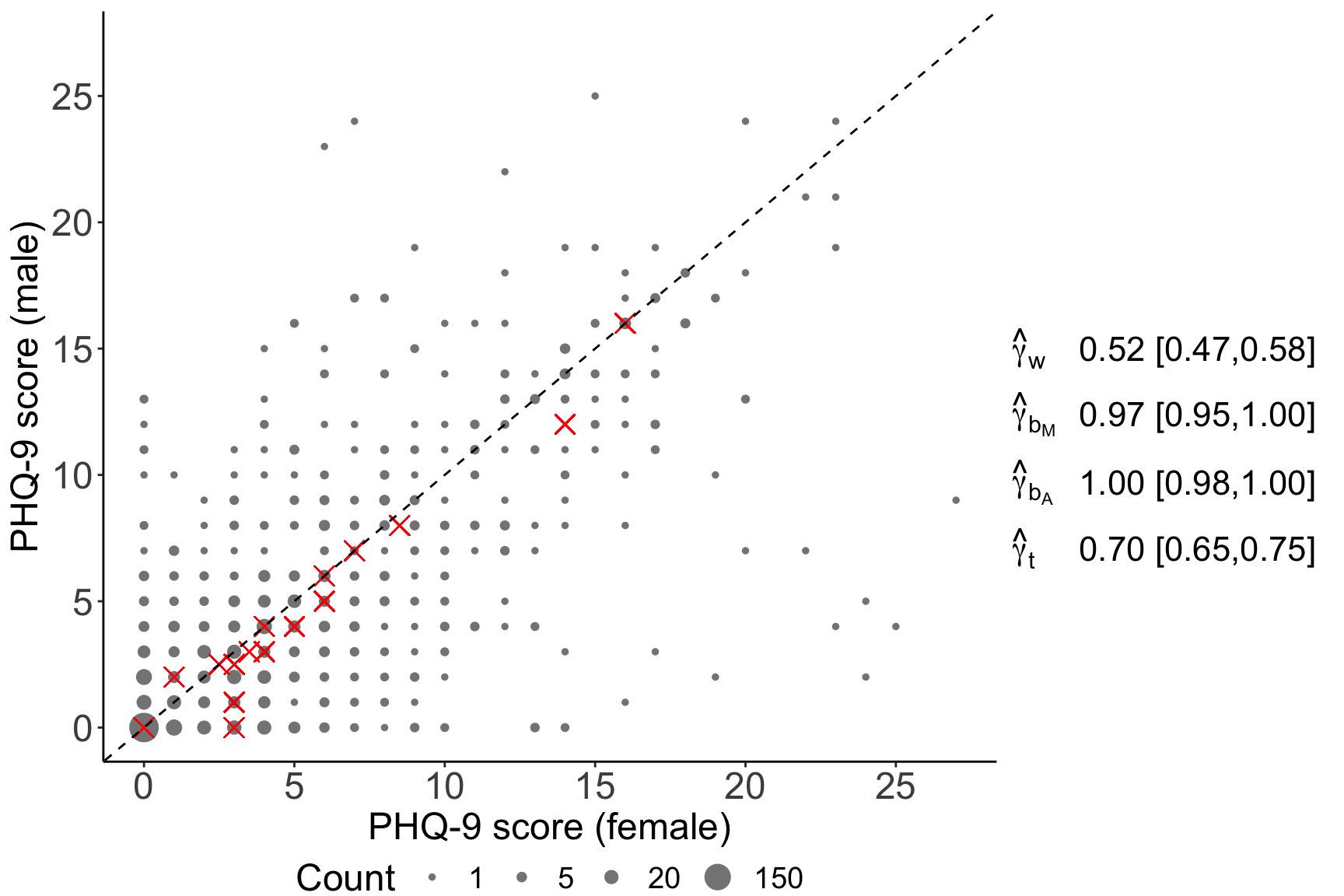}\vspace{-0.1cm}
         \caption{PHQ-9 scores}\vspace{0.3cm}
         \label{fig:phq9}
     \end{subfigure}\\
          \begin{subfigure}[b]{0.5\textwidth}
         \centering
         \includegraphics[width=0.9\textwidth]{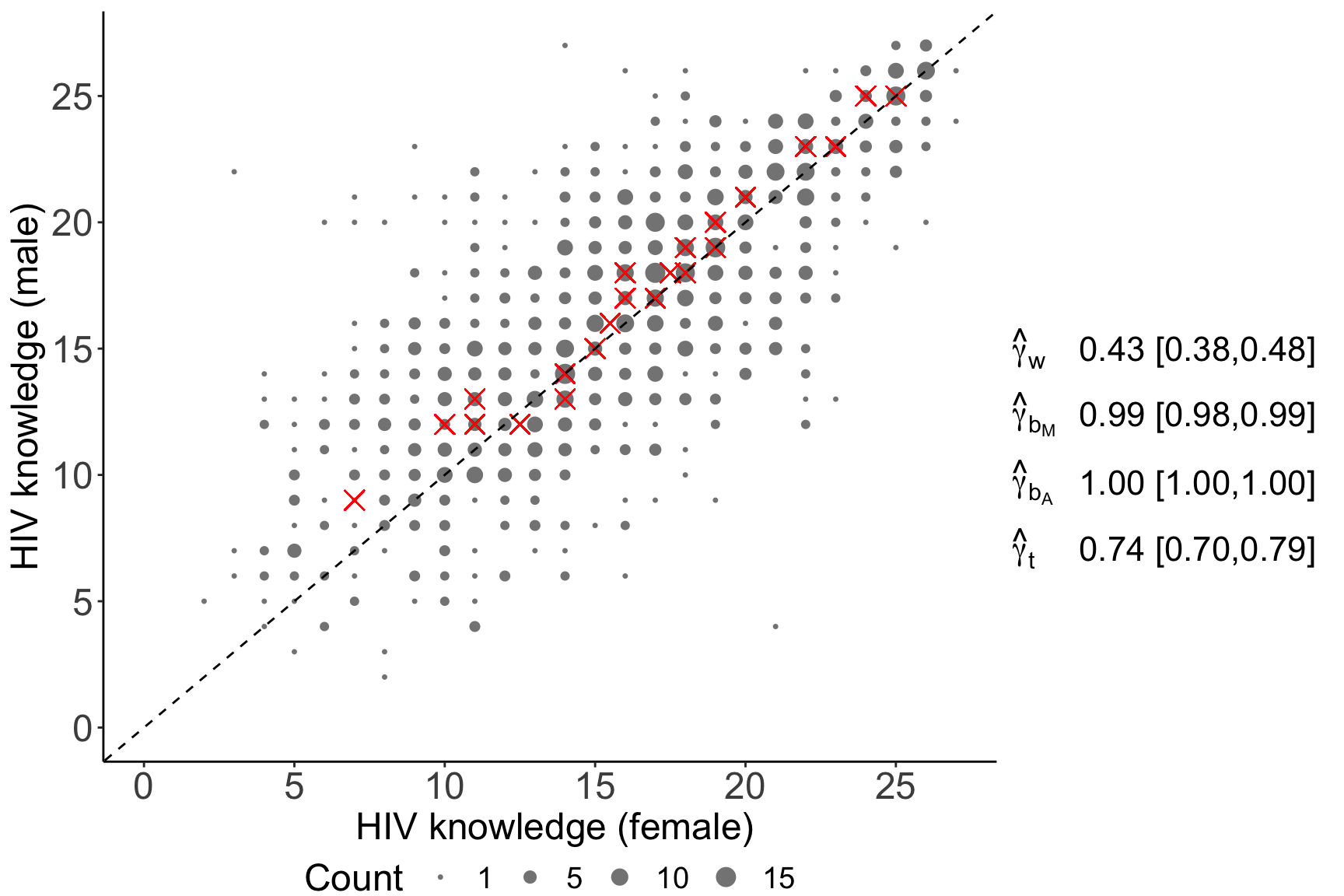}\vspace{-0.1cm}
         \caption{HIV knowledge}\vspace{0.3cm}
         \label{fig:hivknow}
     \end{subfigure}\hfill
     \begin{subfigure}[b]{0.5\textwidth}
         \centering
         \includegraphics[width=0.9\textwidth]{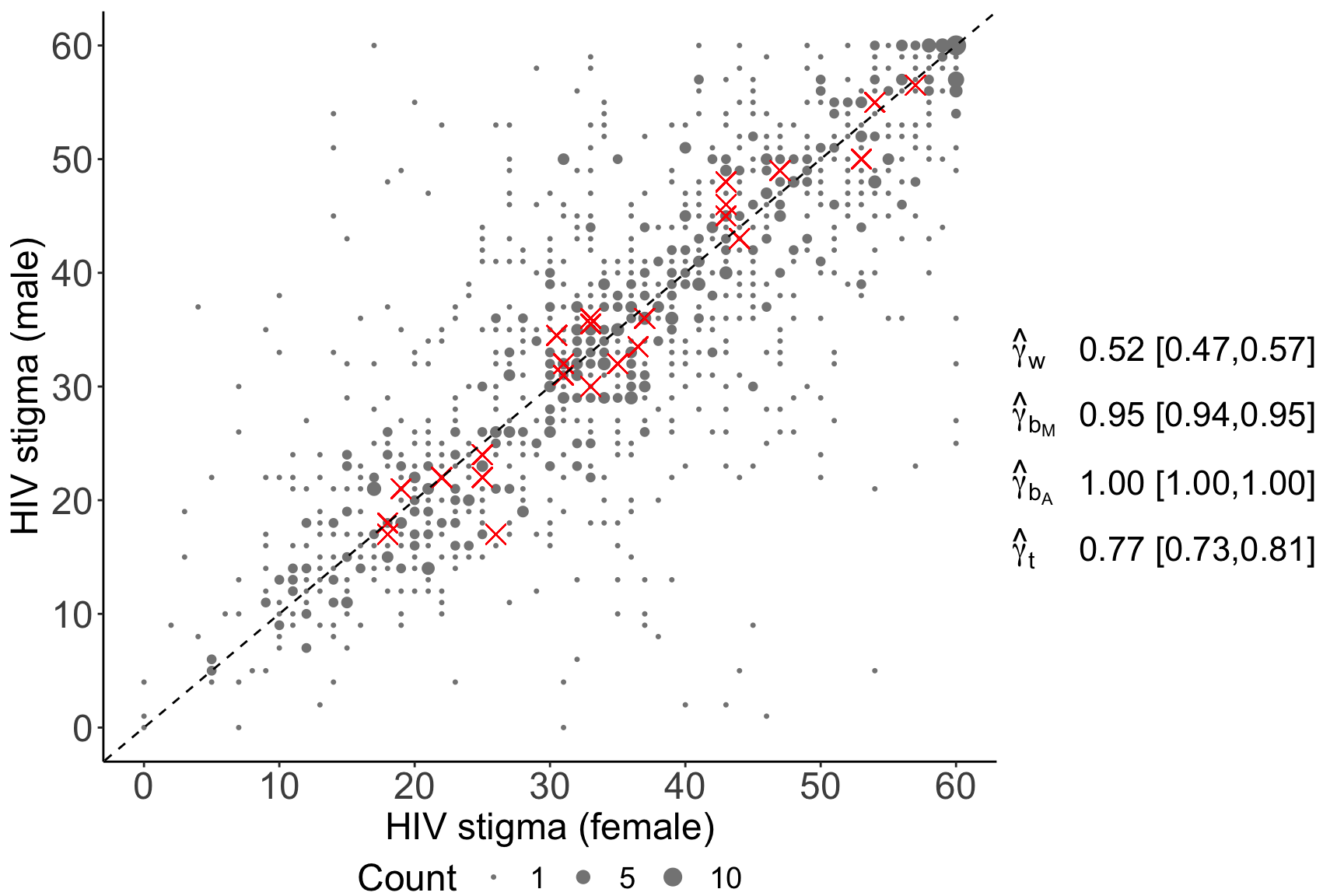}\vspace{-0.1cm}
         \caption{HIV stigma}\vspace{0.3cm}
         \label{fig:hivstigma}
     \end{subfigure}\\
    \begin{subfigure}[b]{0.5\textwidth}
         \centering
         \includegraphics[width=0.9\textwidth]{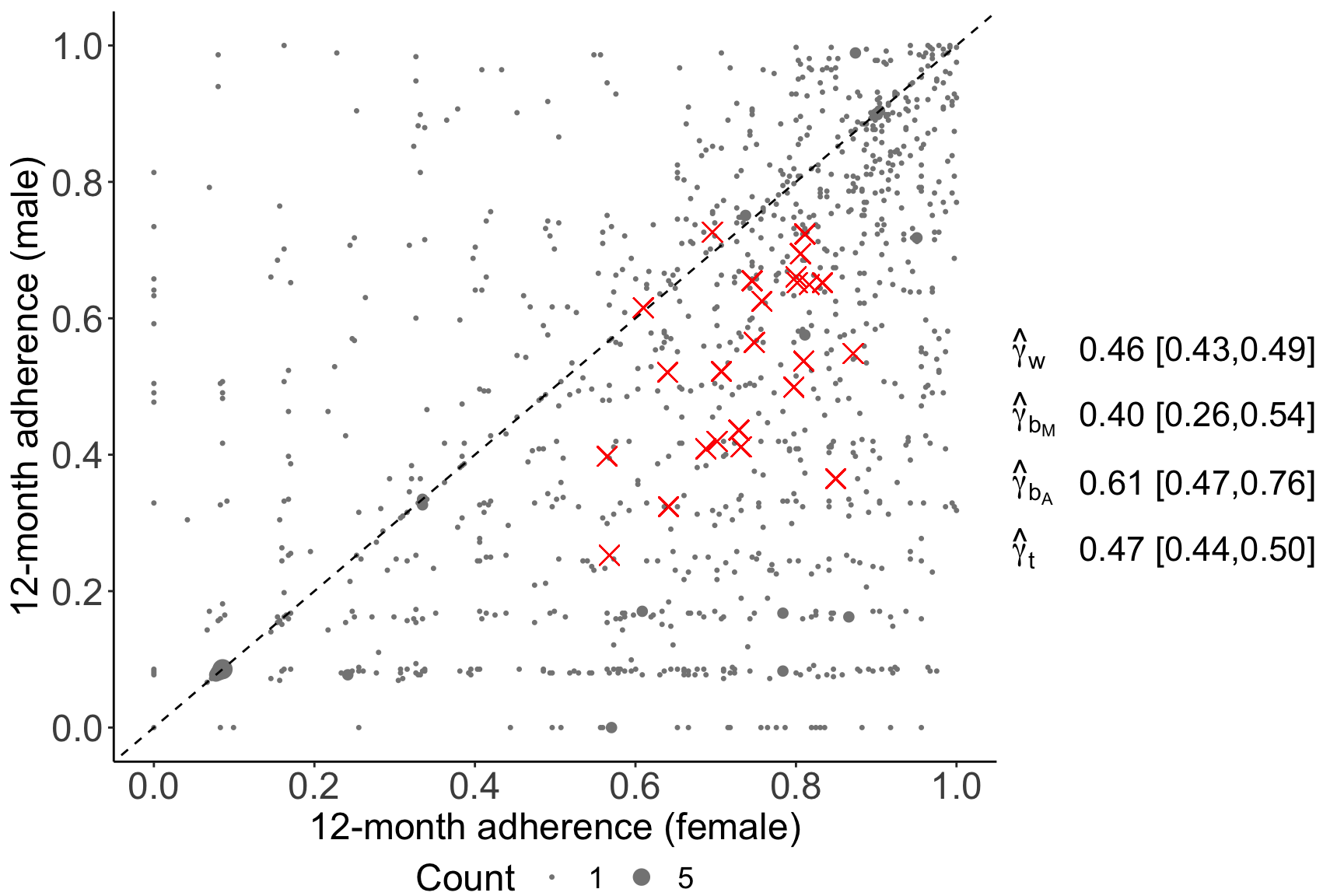}\vspace{-0.1cm}
         \caption{12-month adherence}\vspace{0.3cm}
         \label{fig:adhere}
    \end{subfigure}\hfill
     \caption[]{Scatter plots of PHQ-9 scores, age at enrollment (years), HIV knowledge, HIV stigma, and 12-month adherence (\%) of female and male partners enrolled in the clustered randomized clinical trial. The red cross sign represents the sample cluster median and the dot sign represents the observation. The right side of each subfigure shows the estimates (95\% confidence intervals) of within-cluster, between-cluster, and total Spearman rank correlations ($\hat{\gamma}_w$, $\hat{\gamma}_{b_{M}}$, $\hat{\gamma}_{b_{A}}$, $\hat{\gamma}_t$), where $\hat{\gamma}_{b_{M}}$ is the estimator of $\gamma_b$ based on cluster medians obtained from CPM and $\hat{\gamma}_{b_A}$ is the estimator of $\gamma_b$ based on the linear approximation. }
        \label{fig:example2}
\end{figure}

The Homens para $\text{Sa}\acute{\text{u}}\text{de}$ Mais (HoPS+) study is a cluster randomized controlled trial in $\text{Zamb}\acute{\text{e}}\text{zia}$ Province, Mozambique \citep{audet2018, Audet2024}. The trial was designed to measure the impact of incorporating male partners with HIV into prenatal care for pregnant women living with HIV on adherence to treatment. The trial enrolled 1073 participating couples living with HIV at 24 clinical sites. The number of couples at clinical sites ranged from 15 to 71. At the time of randomization (baseline), age, depressive symptoms measured by Patient Health Questionnaire-9 (PHQ-9) score, HIV knowledge, and HIV stigma were captured. We are interested in the correlation of these baseline measures within couples. We are also interested in the correlation of 12-month adherence to ART within couples. Figure \ref{fig:example2} shows scatter plots of these measures. Some of these measures are ordinal (e.g., PHQ-9) and some are skewed (e.g., adherence).

In this example, the cluster is the clinical site and the observations are made on couples (e.g., $X$ = age of female partners, $Y$ = age of male partners). Hence, it is reasonable to assign equal weights to couples. The estimates of the within-cluster, between-cluster, and total Spearman rank correlations are shown in Figure \ref{fig:example2}. The total Spearman rank correlation for age, 0.59, was moderate to strong. The between-cluster Spearman rank correlation for age was $\hat \gamma_{b_M} =0.38$, suggesting weak to moderate correlation between median male and female ages within clinical sites. The within-cluster Spearman rank correlation was 0.61, implying that after controlling for clinical site, the correlation between couples for age remained high. For PHQ-9 scores, HIV knowledge, and HIV stigma, the total Spearman rank correlation between couples was strong, ranging from 0.70 to 0.77. The correlation became moderate after controlling for clinical sites, with $\hat \gamma_{w}$ varying from 0.43 to 0.52. The between-cluster Spearman rank correlations for the three measures were extremely strong, which can be seen in Figure \ref{fig:example2}. The approximation-based estimates ($\hat{\gamma}_{b_A}$) of the between-cluster correlation hit the boundary and were thus set to be 1, and the cluster-median-based estimates ($\hat{\gamma}_{b_M}$) were close to 1. Taken as a whole, these estimates suggest that the scores between male and female partners for these measures are highly correlated but that some of the correlation is due to similarities within sites. This may reflect differences between participants across sites or perhaps differences in the ways the questionnaires were administrated across study sites. Finally, the total correlation for 12-month adherence was moderate, 0.47. After controlling for clinical sites, the correlation remained moderate, $\hat \gamma_{w}=0.46$. The between-cluster correlation was moderate to high, $\hat \gamma_{b_M}=0.40$ and $\hat \gamma_{b_A}=0.61$; this difference might be due to the small intraclass correlation for this variable (the rank ICC of 12-month adherence for males was 0.06 and for females was 0.07).

\clearpage

\section{Discussion}
In this paper, we defined the population parameters of the between- and within-cluster Spearman rank correlations for clustered data, which are natural extensions of the between- and within-cluster Pearson correlations to the rank scale. We also approximated their relationship with the total Spearman rank correlation and the rank intraclass correlation coefficient. Compared with the traditional Pearson correlation, our method is insensitive to extreme values and skewed distributions, and does not depend on the scale of the data, providing a robust framework without considering data transformation. Our method is general, and is applicable to any orderable variables without any need to assign arbitrary numbers to ordered categories. Our estimators are asymptotically normal, with generally low bias and good coverage in our simulations. We have developed an R package, \texttt{rankCorr}, available on CRAN, which implements our new method \citep{tuRankCorr}.

Our method has some limitations. Our method requires fitting models for the conditional distributions of $X$ and $Y$ given cluster index, which need to be approximately correct to get unbiased estimates. We suggest using semiparametric cumulative probability models which maintain the rank-based nature of Spearman rank correlation. In addition, our estimators of the between-cluster Spearman rank correlation have bias when there is a small number of clusters or when the between- and within-cluster Spearman rank correlations are in opposite directions. The two problems also exist when estimating the between-cluster Pearson correlation. As the cluster size increases, the latter problem goes away.

In practice, one may be interested in estimating covariate-adjusted rank correlations. For example, in the application of CD4 and CD8 data, there may be interest in measuring the rank correlations after adjusting for age. The methods in this manuscript could be extended to allow for covariate adjustment by fitting CPMs that include the covariate, in addition to the cluster indicators. We suspect that the correlation between probability-scale residuals from these fitted models could be used to estimate covariate-adjusted within-cluster rank correlations and that the correlation between cluster indicator coefficients from these fitted models could be used to estimate covariate-adjusted between-cluster rank correlations. This approach is somewhat similar to random effects approaches used for estimating covariate-adjusted within- and between-cluster Pearson correlations \citep{ferrari2005}. Such an approach, as well as Spearman rank correlation as a function of time with longitudinal data, warrants further investigation. 

\section*{Acknowledgments}
We would like to thank the study investigators for providing data used in our example applications. This study was supported in part by funding from the National Institutes of Health (R01AI093234; P30AI110527 and K23AI120875 for the longitudinal biomarker data; and R01MH113478 for cluster randomized controlled trial data).

\bibliographystyle{biom} 
\bibliography{reference1.bib}

\begin{thebibliography}{}

\bibitem[\protect\citeauthoryear{Arakelian and Karlis}{Arakelian and Karlis}{2014}]{arakelian2014}
Arakelian, V. and Karlis, D. (2014).
\newblock Clustering dependencies via mixtures of copulas.
\newblock {\em Commun Stat Simul Comput} {\bf 43,} 1644--1661.

\bibitem[\protect\citeauthoryear{Audet, Graves, Barreto, De~Schacht, Gong, Shepherd, et~al\mbox{.}}{Audet et~al.}{2018}]{audet2018}
Audet, C.~M., Graves, E., Barreto, E., De~Schacht, C., Gong, W., Shepherd, B.~E., et~al. (2018).
\newblock Partners-based {HIV} treatment for seroconcordant couples attending antenatal and postnatal care in rural mozambique: A cluster randomized trial protocol.
\newblock {\em Contemp Clin Trials Commun} {\bf 71,} 63--69.

\bibitem[\protect\citeauthoryear{Audet, Graves, Shepherd, Prigmore, Brooks, Emílio, Matino, Paulo, Diemer, Frisby, Sack, Aboobacar, Barreto, Van~Rompaey, and De~Schacht}{Audet et~al.}{2024}]{Audet2024}
Audet, C.~M., Graves, E., Shepherd, B.~E., Prigmore, H.~L., Brooks, H.~L., Emílio, A., Matino, A., Paulo, P., Diemer, M.~A., Frisby, M., Sack, D.~E., Aboobacar, A., Barreto, E., Van~Rompaey, S., and De~Schacht, C. (2024).
\newblock Partner-based hiv treatment for seroconcordant couples attending antenatal and postnatal care in rural mozambique: A cluster randomized controlled trial.
\newblock {\em J Acquir Immune Defic Syndr} {\bf 96,} 259--269.

\bibitem[\protect\citeauthoryear{Bross}{Bross}{1958}]{Bross1958}
Bross, I. D.~J. (1958).
\newblock How to use ridit analysis.
\newblock {\em Biometrics} {\bf 14,} 18--38.

\bibitem[\protect\citeauthoryear{Castilho, Shepherd, Koethe, Turner, Bebawy, Logan, Rogers, Raffanti, and Sterling}{Castilho et~al.}{2016}]{castilho2016}
Castilho, J., Shepherd, B., Koethe, J., Turner, M., Bebawy, S., Logan, J., Rogers, W., Raffanti, S., and Sterling, T. (2016).
\newblock Cd4+/cd8+ ratio, age, and risk of serious noncommunicable diseases in hiv-infected adults on antiretroviral therapy.
\newblock {\em AIDS} {\bf 30(6),} 899--908.

\bibitem[\protect\citeauthoryear{Ferrari, Al-Delaimy, Slimani, Boshuizen, Roddam, Orfanos, Skeie, Rodríguez-Barranco, Thiebaut, Johansson, Palli, Boeing, Overvad, and Riboli}{Ferrari et~al.}{2005}]{ferrari2005}
Ferrari, P., Al-Delaimy, W.~K., Slimani, N., Boshuizen, H.~C., Roddam, A., Orfanos, P., Skeie, G., Rodríguez-Barranco, M., Thiebaut, A., Johansson, G., Palli, D., Boeing, H., Overvad, K., and Riboli, E. (2005).
\newblock {An Approach to Estimate Between- and Within-Group Correlation Coefficients in Multicenter Studies: Plasma Carotenoids as Biomarkers of Intake of Fruits and Vegetables}.
\newblock {\em Am J Epidemiol} {\bf 162,} 591--598.

\bibitem[\protect\citeauthoryear{Fisher}{Fisher}{1925}]{fisher1925}
Fisher, R. (1925).
\newblock {\em Statistical methods for research workers.}
\newblock Edinburgh Oliver \& Boyd.

\bibitem[\protect\citeauthoryear{Genest and Ne\v{s}lehov\'{a}}{Genest and Ne\v{s}lehov\'{a}}{2007}]{genest2007}
Genest, C. and Ne\v{s}lehov\'{a}, J. (2007).
\newblock A primer on copulas for count data.
\newblock {\em ASTIN Bulletin} {\bf 37,} 475--515.

\bibitem[\protect\citeauthoryear{Harrell}{Harrell}{2016}]{harrell2015}
Harrell, F.~E. (2016).
\newblock {\em rms: Regression Modeling Strategies}.
\newblock R package version 4.5–0.

\bibitem[\protect\citeauthoryear{Hunsberger, Long, Reese, Hong, Myles, Zerbe, Chetchotisakd, and Shih}{Hunsberger et~al.}{2022}]{hunsberger2022}
Hunsberger, S., Long, L., Reese, S.~E., Hong, G.~H., Myles, I.~A., Zerbe, C.~S., Chetchotisakd, P., and Shih, J.~H. (2022).
\newblock Rank correlation inferences for clustered data with small sample size.
\newblock {\em Stat Neerl} {\bf 76,} 309--330.

\bibitem[\protect\citeauthoryear{Kendall}{Kendall}{1938}]{kendall1938}
Kendall, M.~G. (1938).
\newblock A new measure of rank correlation.
\newblock {\em Biometrika} {\bf 30,} 81--89.

\bibitem[\protect\citeauthoryear{Kendall}{Kendall}{1970}]{kendall1970}
Kendall, M.~G. (1970).
\newblock {\em Rank Correlation Methods}.
\newblock Griffin, London.

\bibitem[\protect\citeauthoryear{Kosmidis and Karlis}{Kosmidis and Karlis}{2016}]{kosmidis2016}
Kosmidis, I. and Karlis, D. (2016).
\newblock Model-based clustering using copulas with applications.
\newblock {\em Stat Comput} {\bf 26,} 1079--1099.

\bibitem[\protect\citeauthoryear{Kruskal}{Kruskal}{1958}]{kruskal1958}
Kruskal, W.~H. (1958).
\newblock Ordinal measures of association.
\newblock {\em J Am Stat Assoc} {\bf 53,} 814--861.

\bibitem[\protect\citeauthoryear{Li and Shepherd}{Li and Shepherd}{2012}]{li2012}
Li, C. and Shepherd, B.~E. (2012).
\newblock {A new residual for ordinal outcomes}.
\newblock {\em Biometrika} {\bf 99,} 473--480.

\bibitem[\protect\citeauthoryear{Liu, Li, Wanga, and Shepherd}{Liu et~al.}{2018}]{liu2018}
Liu, Q., Li, C., Wanga, V., and Shepherd, B.~E. (2018).
\newblock Covariate-adjusted spearman's rank correlation with probability-scale residuals.
\newblock {\em Biometrics} {\bf 74,} 595--605.

\bibitem[\protect\citeauthoryear{Liu, Shepherd, Li, and Harrell~Jr.}{Liu et~al.}{2017}]{liu2017}
Liu, Q., Shepherd, B.~E., Li, C., and Harrell~Jr., F.~E. (2017).
\newblock Modeling continuous response variables using ordinal regression.
\newblock {\em Stat Med} {\bf 36,} 4316--4335.

\bibitem[\protect\citeauthoryear{Nelsen}{Nelsen}{2006}]{nelsen2006}
Nelsen, R.~B. (2006).
\newblock {\em An Introduction to Copulas}.
\newblock Springer, New York, 2nd edition.

\bibitem[\protect\citeauthoryear{Pearson}{Pearson}{1907}]{Pearson1907}
Pearson, K.~G. (1907).
\newblock {\em On further methods of determining correlation}.
\newblock Cambridge University Press.

\bibitem[\protect\citeauthoryear{Rosner and Glynn}{Rosner and Glynn}{2017}]{rosner2017}
Rosner, B. and Glynn, R.~J. (2017).
\newblock Estimation of rank correlation for clustered data.
\newblock {\em Stat Med} {\bf 36,} 2163--2186.

\bibitem[\protect\citeauthoryear{Shepherd, Li, and Liu}{Shepherd et~al.}{2016}]{shepherd2016}
Shepherd, B., Li, C., and Liu, Q. (2016).
\newblock Probability-scale residuals for continuous, discrete, and censored data.
\newblock {\em Can J Stat} {\bf 44,} 463--479.

\bibitem[\protect\citeauthoryear{Shih and Fay}{Shih and Fay}{2017}]{shih2017}
Shih, J.~H. and Fay, M.~P. (2017).
\newblock Pearson's chi-square test and rank correlation inferences for clustered data.
\newblock {\em Biometrics} {\bf 73,} 822--834.

\bibitem[\protect\citeauthoryear{Snijders and Bosker}{Snijders and Bosker}{1999}]{snijders1999}
Snijders, T. and Bosker, R. (1999).
\newblock {\em Multilevel Analysis: An Introduction to Basic and Advanced Multilevel Modeling}.
\newblock Sage Publishers, London.

\bibitem[\protect\citeauthoryear{Spearman}{Spearman}{1904}]{spearman1904}
Spearman, C. (1904).
\newblock The proof and measurement of association between two things.
\newblock {\em J Psychopathol Clin Sci} {\bf 15,} 72--101.

\bibitem[\protect\citeauthoryear{Stefanski and Boos}{Stefanski and Boos}{2002}]{stefanski2002}
Stefanski, L.~A. and Boos, D.~D. (2002).
\newblock The calculus of m-estimation.
\newblock {\em Am Stat} {\bf 56,} 29--38.

\bibitem[\protect\citeauthoryear{Tian, Shepherd, Li, Zeng, and Schildcrout}{Tian et~al.}{2023}]{tian2023}
Tian, Y., Shepherd, B.~E., Li, C., Zeng, D., and Schildcrout, J.~S. (2023).
\newblock Analyzing clustered continuous response variables with ordinal regression models.
\newblock {\em Biometrics} {\bf 79,} 3764--3777.

\bibitem[\protect\citeauthoryear{Tu, Li, and Shepherd}{Tu et~al.}{2023}]{tuRankCorr}
Tu, S., Li, C., and Shepherd, B.~E. (2023).
\newblock {\em rankCorr: Total, Between-, and Within-Cluster Spearman Rank Correlations for Clustered Data}.
\newblock R package version 4.2–1.

\bibitem[\protect\citeauthoryear{Tu, Li, Zeng, and Shepherd}{Tu et~al.}{2023}]{tu2023}
Tu, S., Li, C., Zeng, D., and Shepherd, B.~E. (2023).
\newblock Rank intraclass correlation for clustered data.
\newblock {\em Stat Med} {\bf 42,} 4333--4348.

\end{thebibliography}

\section*{Supporting information}
Web Appendices and Tables referenced in Sections \ref{inference} and \ref{simulations} may be found in the online version of the article at the publisher’s website. We have developed an R package, \texttt{rankCorr}, available on CRAN, which implements our new method. The codes for simulations and applications are available on GitHub (https://github.com/shengxintu/rankCorr).

\end{document}